\documentclass[twocolumn,dutch,english,amsmath,amssymb,aps,prl,showpacs]{revtex4}
\usepackage[T1]{fontenc}
\usepackage[latin9]{inputenc}
\usepackage{color}
\usepackage{babel}
\usepackage{mathrsfs}
\usepackage{amsmath}
\usepackage{amssymb}
\usepackage{graphicx}
\usepackage{esint}
\usepackage[unicode=true,pdfusetitle,
 bookmarks=false,
 breaklinks=false,pdfborder={0 0 1},backref=false,colorlinks=true]
 {hyperref}
\hypersetup{
 colorlinks,linkcolor=blue,citecolor=blue,urlcolor=blue}

\makeatletter

\newcommand*\LyXThinSpace{\,\hspace{0pt}}

\@ifundefined{textcolor}{}
{%
 \definecolor{BLACK}{gray}{0}
 \definecolor{WHITE}{gray}{1}
 \definecolor{RED}{rgb}{1,0,0}
 \definecolor{GREEN}{rgb}{0,1,0}
 \definecolor{BLUE}{rgb}{0,0,1}
 \definecolor{CYAN}{cmyk}{1,0,0,0}
 \definecolor{MAGENTA}{cmyk}{0,1,0,0}
 \definecolor{YELLOW}{cmyk}{0,0,1,0}
}

\usepackage{babel}
\usepackage{babel}
\usepackage{babel}

\usepackage{babel}
\usepackage{mathrsfs}

\makeatother

\begin{document}

\title{Quantum Diagrammatic Theory of the Extrinsic Spin Hall Effect in
Graphene}

\author{Mirco Milletar\`{i}}

\email{milletari@gmail.com}

\selectlanguage{english}%

\affiliation{Centre for Advanced 2D Materials and Department of Physics, National
University of Singapore, Singapore, 117551}

\author{Aires Ferreira}

\email{aires.ferreira@york.ac.uk}

\selectlanguage{english}%

\affiliation{Department of Physics, University of York, York YO10 5DD, United
Kingdom}
\begin{abstract}
\textcolor{black}{We present a rigorous microscopic theory of the
extrinsic spin Hall effect in disordered graphene based on a }{\textcolor{black}{nonperturbative}}\textcolor{black}{{}
quantum diagrammatic treatment incorporating skew scattering and }\textcolor{black}{\emph{anomalous}}\textcolor{black}{---impurity
concentration-independent---quantum corrections on equal footing.
The leading skew scattering contribution to the spin Hall conductivity
is shown to quantitatively agree with Boltzmann transport theory over
a wide range of parameters. Our self-consistent approach---where all
topologically equivalent }\foreignlanguage{dutch}{\textcolor{black}{noncrossing}}\textcolor{black}{{}
diagrams are resummed---unveils that the skewness generated by spin--orbit-active
impurities deeply influences the anomalous component of the spin Hall
conductivity, even in the weak scattering regime. This seemingly counterintuitive
result is explained by the rich sublattice structure of scattering
potentials in graphene, for which traditional Gaussian disorder approximations
fail to capture the intricate correlations between skew scattering
and side jumps generated through diffusion. Finally, we assess the
role of quantum interference corrections by evaluating an important
subclass of crossing diagrams recently considered in the context of
the anomalous Hall effect, the $X$ and $\Psi$ diagrams {[}Ado }\textcolor{black}{\emph{et
al}}\textcolor{black}{., EPL 111, 37004 (2015){]}. We show that $\Psi$
diagrams---encoding quantum coherent skew scattering---display a strong
Fermi energy dependence, dominating the anomalous spin Hall component
away from the Dirac point. Our findings have direct implications for
nonlocal transport experiments in spin--orbit-coupled graphene systems.}
\end{abstract}


\maketitle

\section{Introduction\label{sec:I-Introduction}}

Spin Hall effects\textemdash the collection of transport phenomena
whereby charge currents propagating in nonmagnetic materials are converted
to transverse spin currents and vice versa \cite{SHE_Dyakanov,SHE_Hirsch,SHE_Zhang}\textemdash constitute
a rapidly evolving front of spintronics research. Following their
first demonstrations in semiconductors and metals \cite{SHE_Kato_Wunderlich_04,SHE_Saitoh_Valenzuela_06},
the spin Hall effects (SHEs) have been explored to devise novel schemes
for interconversion of spin and charge signals. In particular, spin--orbit
torques induced by the SHE from heavy metals have been explored to
manipulate the magnetization dynamics in ferromagnet--metal bilayers,
including tuning of spin relaxation and spin-torque switching of the
magnetized layer \cite{FM_bilayers_Liu11,FM_bilayers_Ando_08,FM_bilayers_Miron_Liu_Liu}.
Conversely, the inverse SHE \cite{SHE_Saitoh_Valenzuela_06} enables
the transformation of pure spin currents injected by spin pumping
from precessing ferromagnets into electric signals \cite{FM_bilayers_SpinPumping}.
Spin--orbit interactions are also of paramount importance in the emergent
field of ``spin caloritronics'', where the inverse SHE is utilized
to detect spin currents generated by the spin Seebeck effect \cite{SpinCaloritronics_Various}.

\begin{figure}[t]
\centering{}\includegraphics[width=0.55\columnwidth]{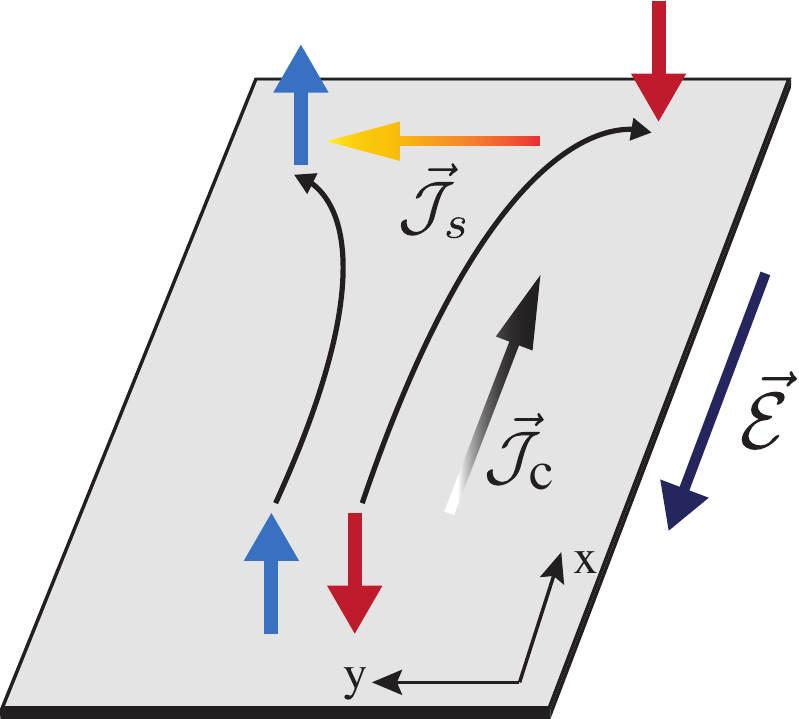}\caption{\label{fig:01_Schematic}Schematic of the SHE in a two-dimensional
material. An external electrical field drives a charge and spin Hall
currents. The reduced dimensionality defines $Oz$ as a preferred
spin direction \cite{SHE_Hirsch}. }
\end{figure}

The SHE efficiency of a material is characterized by the spin Hall
angle, defined as the ratio of $z-$polarized transverse spin current
to longitudinal charge current densities in the steady state, $\gamma=\mathcal{J}_{\perp}/J_{\parallel}$
(see schematic in Fig.~\ref{fig:01_Schematic}). In time-reversal
invariant systems, Onsager reciprocity relations dictate that the
strengths for the direct and inverse SHE are the same, and hence $\gamma$
is an important figure of merit for applications exploring relativistic
spin--orbit coupling transport phenomena. Since the SHEs have their
origin in the coupling between spin and orbital degrees of freedom,
clean metals with large spin--orbit coupling \cite{SHE_4d5dMetals_Tanaka_Wang}
and disordered metals with impurity resonances split by the spin--orbit
interaction \cite{SHE_RS_Metals_Fert_and_Niimi_11} can display robust
SHEs, with $\gamma$ in the range $0.01-0.1$.

The impurity generated extrinsic SHE is of particular interest, from
both applied and fundamental perspectives. In the presence of local
spin--orbit interactions, up and down spin components of wavepackets
are preferably scattered in opposite directions (skew scattering),
leading to the establishment of net spin Hall currents. The degree
of skewness, and thus the resulting spin Hall angles, can be modified
by varying the impurity concentration or by taking different combinations
of host and impurity systems \cite{SHE_RS_Metals_Fert_and_Niimi_11,SHE_Metals_Ingrid}.
This allows to optimize metallic thin films for usage in spin-current
generation (direct SHE) and detection (inverse SHE) schemes.\textcolor{black}{{}
Another appealing scenario is the}\textcolor{black}{\emph{ in situ}}\textcolor{black}{{}
tuning of spin Hall angles for low-power spintronics schemes based
on }\textcolor{black}{\emph{pure}}\textcolor{black}{{} (charge neutral)
spin currents. The latter is a formidable task that requires the ability
to routing pure spin currents by means of external gates. A promising
candidate is the robust extrinsic SHE predicted to occur in graphene
with dilute spin--orbit-active scattering centers \cite{Ferreira14}.
One can envisage that in the vicinity of a }sharp\textcolor{black}{{}
impurity resonance, the spin Hall angle would undergo major changes
upon tuning of the chemical potential, enabling to reverse the sign
of spin currents. The reversible manipulation of charge transport
properties through electrical control of impurity resonances has been
recently reported in dual-gated fluorinated bilayer graphene devices
\cite{RS_Ferreira_Jun15}, suggesting that similar setups could be
explored to achieve gate-tunable spin currents in graphene. }

\textcolor{black}{An interesting} feature of two-dimensional materials
\textcolor{black}{is the possibility to introduce spin--orbit coupling
(SOC) with different symmetries \cite{McCann12,Pachoud} and varying
spatial extent (sub-nanometer range using adatoms \cite{SOC_G_Balakrishnan_13},
nanometer scale using clusters \cite{SOC_G_Balakrishnan_14}, and
spatially uniform SOC through proximity effect to suitable substrates
\cite{SOC_G_Avsar_14,SOC_G_Wang_Morpurgo_15,SOC_G_Calleja15}). The
important role played by the SOC symmetry in the resonant scattering
regime has been elucidated by recent theoretical studies \cite{Pachoud,Cazalilla16}.
The suitability of graphene for all-electrical spintronics is further
supported by recent experimental reports on non-local transport in
adatom-decorated graphene in Refs.~\cite{SOC_G_Balakrishnan_13,SOC_G_Balakrishnan_14}
and, more recently, on spin pumping in graphene from a magnetic insulator
substrate }\textcolor{blue}{\cite{SOC_G_Mendes_15}}\textcolor{black}{.
We finally note that the negligible intrinsic spin--orbit coupling
in the band structure of graphene \cite{Hernando06,Fabian09} is particularly
advantageous, as spin Hall currents generated from impurities can
propagate large distances without suffering from additional spin relaxation
\cite{Ferreira14}. }

These recent developments in graphene spintronics motivate us to further
investigate the microscopic mechanisms underlying the SHE in models
of two-dimensional (2D) massless Dirac fermions subject to spin--orbit
interactions. The giant extrinsic SHE proposed in Ref.~\cite{Ferreira14}
has its origin in the resonant scattering mechanism ubiquitous in
disordered 2D massless Dirac fermions \cite{RS_Stauber,RS_Robinson,RS_Wehling,RS_Ferreira}.
Broadly speaking, the vanishingly small density of states of bare
graphene, $\nu(\epsilon)\propto\epsilon$, favors the appearance of
sharp impurity resonances, and, consequently, large scattering skewness
in the presence of SOC. The full conductivity tensor in the charge
and spin sectors, including charge--spin transverse (Hall) conductivities,
can be conveniently computed by means of Boltzmann transport theory
upon careful identification of the transport lifetimes determining
asymmetric distortions of the Fermi surface induced by the SOC \cite{Ferreira14}.
The simple semiclassical approach is justified well inside the dilute
limit, where skew scattering provides the leading contribution to
the spin Hall conductivity, \emph{i.e.}, $\sigma_{\textrm{SH}}=\mathcal{S}(\epsilon)n_{\textrm{}}^{-1}$,
where $n_{\textrm{}}\ll1$ is the SOC-active impurity concentration,
and $\mathcal{S}(\epsilon)$ is some non-universal function of the
Fermi energy and microscopic parameters of the model. On the other
hand, for not too dilute concentrations (e.g., in the range 0.01--0.1$\%$
atomic ratio for resonant impurities) other extrinsic mechanisms can
generate spin Hall currents in graphene \cite{Milletari_1}. In particular,
the trajectory of charge carriers can undergo a transverse spin-dependent
displacement upon scattering from a spin--orbit-coupled impurity---so-called
quantum side jump---which gives rise to a net contribution to the
spin Hall current. The latter shows up in the next leading term in
the spin Hall conductivity expansion, 
\begin{equation}
\sigma_{\textrm{SH}}=\mathcal{S}(\epsilon)n^{-1}+\mathcal{Q}(\epsilon)+...\:,\label{eq:SH_imp_dens_expansion}
\end{equation}
here referred to as the \emph{anomalous} contribution. The determination
of the quantum side jump contribution to $\mathcal{Q}(\epsilon)$
within semiclassical transport theory has been the cause of much debate.
Historically, this controversy originated in the closely related anomalous
Hall effect (AHE), taking place in ferromagnetic materials~\cite{Synitsin_review}.
The controversy surrounding the semiclassical description of the side-jump
mechanism originated from its association with the Berry connection,
which is a gauge dependent quantity. Recently, a semiclassical formulation
preserving $U(1)$ gauge invariance has been developed by Synitsin
and co-workers \cite{Synitsin_coordinate_shift}, which allows to
treat quantum side jump (QSJ) contributions in the weak scattering
limit. The quantum linear response theory or the quasiclassical Keldysh
approach~\cite{Roberto} provide powerful alternatives to semiclassical
approaches. In this paper, we will use the linear response theory
(LRT), where different contributions to the spin Hall (SH) conductivity
can be evaluated systematically by means of the diagrammatic technique.
Whereas previous use of diagrammatic expansions in studies of the
SHE/AHE have traditionally assumed weak disorder (Gaussian) approximations,
a proper treatment of the skew scattering mechanism in graphene necessarily
requires a nonperturbative approach. The most pressing question is
how to treat semiclassical skew scattering ($\mathcal{S}$) and anomalous
quantum scattering ($\mathcal{Q}$) processes when scattering potentials
are no longer weak\textcolor{black}{{} or exhibit a rich structure,
}e.g., breaking pseudo-spin rotational invariance through a mass term
\cite{Milletari_1}. These questions are of much interest in graphene
systems, where impurities generally have a complex pseudo-spin--valley
texture \cite{Pachoud,Graphene_SublatticeValleyTexture_15}. In addition,
the role of quantum coherent multiple scattering in the SHE remains
nearly a virgin territory.

In this paper, we tackle the aforementioned issues by means of a simple,
yet powerful, extension of the standard diagrammatic approach originally
developed by Baym in the context of the quantum kinetic equation (Kadanoff-Baym)
formalism~\cite{Baym}. We show that a proper evaluation of vertex
corrections allows to take into account skew scattering and quantum
processes at \emph{all orders} in perturbation theory by means of
exact resummations.\textcolor{black}{{} We find that while the single-impurity
(semiclassical) skew scattering contribution }\textcolor{black}{\emph{quantitativel}}\textcolor{black}{y
agrees with Boltzmann theory, the anomalous component of $\sigma_{\textrm{SH}}$
shows a richer structure, with several contributions beyond the semiclassical
QSJ processes. One of our main results concern the role of quantum
interference: coherent skew scattering from two impurities is found
to provide a remarkably large contribution to $\mathcal{Q}$, opening
doors to the observation of quantum coherent processes in non local
transport experiments.}

This paper is organized as follows. In Section \ref{sec:II-METHODOLOGY},
we set the notation and outline the extended LRT formalism employed
in the reminder of the article. In this section, we also comment on
the different types of approximations commonly employed in theoretical
studies of AHE/SHE. Section \ref{sec:III-MODEL SYSTEM} introduces
the disordered spin--orbit-coupled graphene model system under examination,
and Sec.\,\ref{sec:IV-NON_CROSSING_WEAK_SCATT} presents the calculation
of the SH conductivity within the weak scattering regime. The scope
of this section is to highlight the shortcomings of the widely used
white-noise \textcolor{black}{Gaussian disorder assumption}s. \textcolor{black}{In
Sec.\,\ref{sec:V-NONPERTURBATIVE} we compare the simple Gaussian
result with the full $T$ matrix calculation and show that the former
misses an important Fermi energy dependence. }Furthermore, we show
that the (semiclassical) skew scattering contribution within the extended
LRT formalism \emph{quantitatively} agrees with the exact solution
of the corresponding Boltzmann transport equations. Up to our knowledge
this is the first work reporting the equivalence between the two approaches
in the strong scattering regime. Finally, in Sec.\,\ref{sec:VII-CROSSING}
we improve upon the noncrossing approximation to incorporate important
coherent multiple scattering contributions. The latter is motivated
by recent findings on the importance of a subclass of crossing diagrams
for a correct description of the AHE in 2D systems of massive Dirac
fermions~\cite{Ado}. By comparing the Gaussian result with the extended
LRT based on the $T$~Matrix approach, we will show that the former
leads to an incorrect quantitative (and qualitative) picture. More
specifically, we show that the Gaussian approximation erroneously
predicts the vanishing of a specific set of crossing diagrams (``$\Psi$''
diagrams). Indeed, within the $T$~matrix approach, we show that
these diagrams give a dominant contribution in some region of parameter
space, demonstrating again the limitations of the Gaussian approximation.

\section{METHODOLOGY\label{sec:II-METHODOLOGY}}

In this paper we are interested in models of disordered graphene where
the out-of-plane spin polarization is conserved (see Sec.\,\ref{sec:III-MODEL SYSTEM}).
In such models, the charge- and spin-current density operators are
given by the standard expressions \cite{ProperDef_Spin} 
\begin{align}
\mathbf{J} & =-e\,\Psi^{\dagger}(\mathbf{x})\,\mathbf{v}\,\Psi(\mathbf{x})\,,\label{eq:charge_current}\\
\boldsymbol{\mathcal{J}} & =-e\,\Psi^{\dagger}(\mathbf{x})\,\frac{1}{2}\{s_{z},\mathbf{v}\}\,\Psi(\mathbf{x})\,,\label{eq:spin_current}
\end{align}
respectively. Here, $\{,\}$ denotes the anti-commutator, $\mathbf{v}$
is the velocity operator, $-e<0$ is the electron's charge, and $s_{z}\equiv s_{3}$
is the diagonal Pauli matrix with eigenvalues $\pm1$. The spin--charge
conductivity tensor describing the interconversion of charge and spin
currents in the presence of SOC is given by the Kubo--Streda formula
\cite{Bruno_01} 
\begin{align}
 & \sigma_{ij}^{z}=\frac{\hbar}{2\pi\Omega}\textrm{Tr}\left\langle J_{i}(G^{R}-G^{A})\mathcal{J}_{j}G^{A}-\right.\nonumber \\
 & \qquad\qquad\qquad\left.\mathcal{J}_{j}(G^{R}-G^{A})J_{i}G^{R}\right\rangle _{\textrm{dis}}\,,\label{eq:SH_cond_def}
\end{align}
where 
\begin{equation}
G^{R(A)}=\frac{1}{\epsilon-H_{0}-V\pm i0^{+}}\,\label{eq:Green_Function_Total}
\end{equation}
is the retarded (advanced) Green function associated with the total
Hamiltonian, $H=H_{0}+V$, with $H_{0}$ denoting the bare term and
$V$ the disorder potential from impurities located at random positions
$\{\mathbf{x}_{i}\}$. The terms $J_{i}$ and $\mathcal{J}_{j}$ are
Cartesian components of the charge- and spin-current density operator,
respectively, and $\langle...\rangle_{\textrm{dis}}$ denotes disorder
configurational average according to the standard prescription 
\begin{equation}
\langle O\rangle_{\textrm{dis}}=\left.\lim_{N,\Omega\rightarrow\infty}\left(\prod_{i=1}^{N}\int_{\Omega}\frac{d^{2}\mathbf{x}_{i}}{\Omega}\right)\,O(\mathbf{x}_{1},...,\mathbf{x}_{N})\,\right|_{\frac{N}{\Omega}=n},\label{eq:conf_average}
\end{equation}
where $i$ labels the impurities and $\Omega$ is the area of the
sample. Finally, $\textrm{Tr}$ denotes the trace over the complete
Hilbert space.

\begin{figure}[!t]
\includegraphics[width=1\columnwidth]{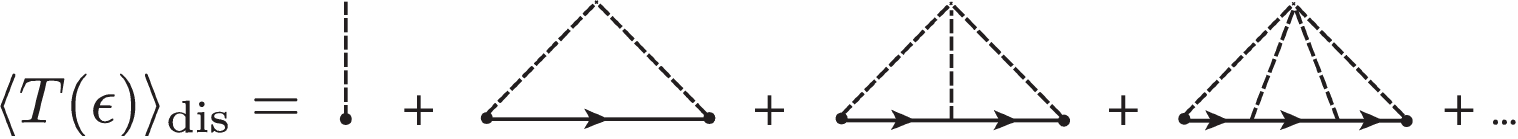} \caption{\label{fig:TMatrix}$T$~Matrix expansion. The truncated series of
diagrams constituting the $T$~matrix is shown up to fourth order
in the impurity potential (black dots). The continuum lines are bare
propagators.}
\end{figure}

\begin{figure*}[th]
\centering{} \includegraphics[width=2\columnwidth]{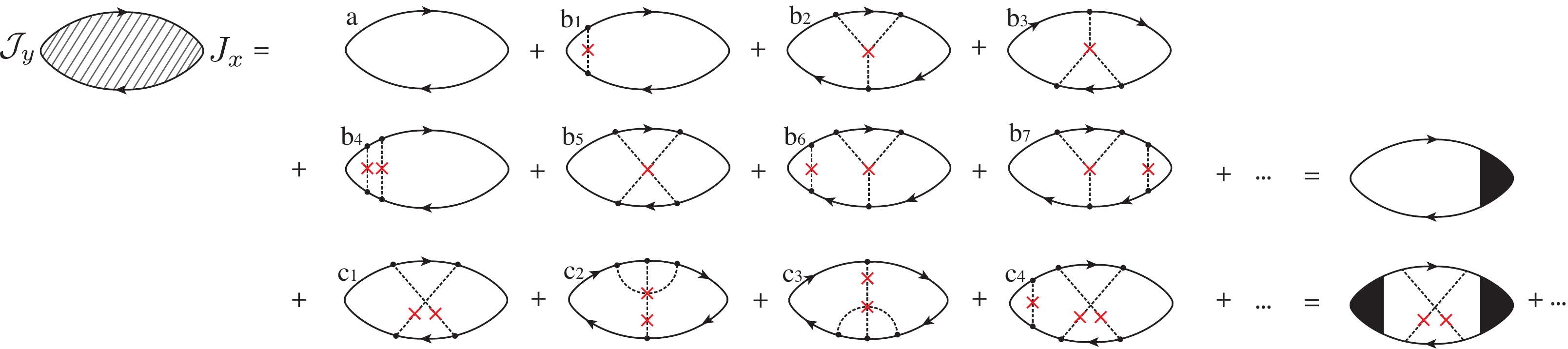} \caption{\label{fig:general_exp}Feynman diagrams considered in this work.
(a) is the empty bubble, and remainder diagrams encode vertex corrections.
The set $\{b_{p}\}_{p=1...\infty}$ is the complete series of noncrossing
2-particle diagrams, which contain the conventional ladder diagrams
as a subset $\{b_{1},b_{4},...\}$. The last line shows crossing diagrams
obtained when the full $T$ matrix ($\mathcal{T}^{\lambda}$) is used
{[}Eq.~(\ref{eq:sH_cond_Baym}){]}. Legend: thick black lines represented
disorder averaged propagators, the circles at the vertices are current
operators, dashed black lines represent potential $W$ insertions,
and the crosses represent impurity density insertions.}
\end{figure*}

Owing to time-reversal symmetry, the off diagonal entries of the tensor
$\sigma_{ij}^{z}$ ($i\neq j$) characterize both the transverse spin
current density generated by a longitudinal electric field, $\mathcal{J}_{i}=\sigma_{ij}^{z}E_{j}$,
and the transverse charge current density induced by a spin-dependent
chemical potential gradient, $J_{j}=\sigma_{ij}^{z}\mathcal{E}_{j}^{z}$.
Note that the charge--spin tensor is antisymmetric with respect of
exchange of direction indexes, $\sigma_{ij}^{z}=-\sigma_{ji}^{z}$.
In what follows, we define $\sigma_{\textrm{SH}}\equiv\sigma_{yx}^{z}$.
We also make use of natural units $\hbar\equiv1\equiv e$ (the units
are restored in the final expressions for the SH conductivity).

The terms in Eq.~\eqref{eq:SH_cond_def} involving Green functions
in the same sector, only contribute at order $n$ and therefore do
not contribute to the processes described in Eq.~\eqref{eq:SH_imp_dens_expansion}
(see Sec.~\ref{sec:IV-NON_CROSSING_WEAK_SCATT} for details). The
SH conductivity then reduces to 
\begin{equation}
\sigma_{\textrm{SH}}=\frac{1}{\pi\Omega}\textrm{Tr}\,\left[\left\langle G^{R}J_{x}G^{A}\right\rangle _{\textrm{dis}}\mathcal{J}_{y}\right]\,,\label{eq:SH_cond_simplified}
\end{equation}
where we used the fact that in Dirac theories, extrinsic SOC does
not generate additional terms in the velocity operator. This allows
us to bring one of the current operators outside the configurational
average. The SH conductivity {[}Eq.~(\ref{eq:SH_cond_simplified}){]}
can be evaluated by means of standard diagrammatic techniques for
disordered electrons~\cite{Rammer}. Studies of the SHE traditionally
evaluate a selection of low-order diagrams encoding scattering processes
in the weak perturbation regime. As explained in the introduction,
the latter approach is not generally suitable for 2D massless Dirac
fermions, where perturbations induced by impurities can be quite strong.
For this reason, we follow the approach originally developed by Baym~\cite{Baym}
in the quantum kinetic equation formalism, which enables a self-consistent
evaluation of the conductivity tensor at the full $T$ matrix level\textcolor{black}{.
The extended formalism has been applied in Ref.~\cite{Hirsch} to
study resonant impurity scattering in anisotropic superconductors.
}In this approach, one introduces disorder averaged Green functions
\begin{equation}
\mathcal{G}^{\lambda}=\frac{1}{\left(G_{0}^{\lambda}\right)^{-1}-\Sigma^{\lambda}},\label{eq:dis_av_Green function}
\end{equation}
where $G_{0}^{\lambda}$ denotes the Green function of the unperturbed
system, $\lambda=\{R,A\}$ specifies retarded/advanced sector and
$\Sigma^{\lambda}=\langle V+VG_{0}^{\lambda}V+...\rangle_{\textrm{dis}}$
is the self energy. Assuming a short range potential of the form $V=\sum_{i}W(\mathbf{x}-\mathbf{x}_{i})$,
the latter can be written as 
\begin{equation}
\Sigma^{\lambda}=n\,T^{\lambda}+\delta\Sigma^{\lambda}\,,\label{eq:self_energy}
\end{equation}
where \textcolor{black}{{} 
\begin{equation}
{\color{black}T^{\lambda}=W\frac{1}{1-G_{0}^{\lambda}W}\,,}\label{eq:Tmatrix_single}
\end{equation}
is the average $T$~matrix describing scattering off a single impurity},
and $\delta\Sigma^{\lambda}$ contains $O(n^{2})$ terms. Here, $\delta\Sigma^{\lambda}$
contains two physically different class of diagrams that are higher
order in the impurity density $n$: those with crossing impurity lines
and those without. The former describe correlated scattering processes
off multiple impurities, while the latter describe uncorrelated processes
taking place at higher impurity density. For this reason, terms without
crossing impurity lines can be easily included in the $T$~matrix
in a self consistent way~\cite{Rammer}. On the other hand, crossing
diagrams are not in general easy to resum; however, these diagrams
are associated with an extra factor of smallness of the order $(k_{F}\,l)^{-1}$,
where $k_{F}$ is the Fermi momentum and $l$ is the mean free path.
Under disorder average and neglecting crossing diagrams in the self
energy, one can recast Eq.~\eqref{eq:SH_cond_simplified} into the
convenient form 
\begin{align}
\sigma_{\textrm{SH}} & =\frac{1}{\pi\Omega}\textrm{Tr}\left[\mathcal{G}^{R}\,J_{x}\,\mathcal{G}^{A}\,\mathcal{J}_{y}\right]\nonumber \\
 & \,+\frac{1}{\pi\Omega}\textrm{Tr}\left[\mathcal{G}^{A}\mathcal{J}_{y}\mathcal{G}^{R}\left\langle \mathcal{T}^{R}\mathcal{G}^{R}J_{x}\mathcal{G}^{A}\mathcal{T}^{A}\right\rangle _{\textrm{dis}}\right],\label{eq:sH_cond_Baym}
\end{align}
with $\mathcal{T}^{\lambda}$ defined by the relation $\mathcal{T}^{\lambda}=V+V\,\mathcal{G}^{\lambda}\,\mathcal{T}^{\lambda}$.
The first line provides the ``empty bubble'' contribution to the
SH conductivity, \emph{i.e.}, diagram $a$ of Fig.~\eqref{fig:general_exp},
and the second line describes the so-called vertex corrections. While
the self energy dresses the bare propagator (2-point function) in
Eq.~\eqref{eq:dis_av_Green function}, the second line of Eq.~\eqref{eq:sH_cond_Baym}
encodes the dressing of the response function (4-point function),
represented diagrammatically by two-particle noncrossing and crossing
diagrams, $b_{i}$ and $c_{i}$, respectively, in Fig.~\ref{fig:general_exp}.

The two-particle noncrossing diagrams ($b_{i}$) contain information
about the standard semiclassical skew scattering and QSJ: two parametrically
distinguishable contributions with origin in incoherent (single-impurity)
scattering. Formally, the resummation of this class of diagrams is
performed substituting $\mathcal{T}^{\lambda}$ in Eq.~(\ref{eq:sH_cond_Baym})
by its disorder average, that is, $\mathcal{T}^{\lambda}\rightarrow T^{\lambda}$.
It is worth noting that the Born approximation $T^{\lambda}\approx V$
leads to the familiar ladder diagrams $b_{1}$, $b_{4}$ ... in Fig.~\ref{fig:general_exp}.
The resummation of the Born ladder series yields the commonly employed
approximation to the QSJ \cite{Bruno_01,Synitsin_link,Nunner07,Berg11,Barnas12}.
However, by keeping the full $T$ matrix structure one effectively
resum all topologically equivalent, two-particle noncrossing diagrams
at all orders in $V$. The additional terms generated by the $T$
matrix insertions encompass not only skew scattering from arbitrarily
strong potentials, but also important corrections to the anomalous
term in Eq.~\eqref{eq:SH_imp_dens_expansion}---\textcolor{black}{see
Secs.~\ref{sec:IV-NON_CROSSING_WEAK_SCATT} and \ref{sub:VI-A-NonCroossing_Full_Diagramm}
for details.}

\textcolor{black}{Finally, the two-particle crossing diagrams ($c_{i}$)
contain quantum corrections arising from coherent multiple scattering
from two or more impurities. Similarly to weak localization corrections
to the longitudinal conductivity, these diagrams come with an extra
factor of smallness $(k_{F}\,l)^{-1}$ due to the crossing of lines
belonging to different impurity density insertions. For this reason,
it was believed that their effect would be relevant only in the deep
quantum regime, $k_{F}\,l\ll1$. However, this argument is not generally
correct, as $c_{i}$ diagrams with two impurity crossing lines also
contribute to order $(k_{F}\,l)^{0}$ in the transverse conductivity,
and therefore correct $\mathcal{Q}(\epsilon)$ in Eq.~\eqref{eq:SH_imp_dens_expansion}.
This was recently discovered by Ado }\textcolor{black}{\emph{et al.}}\textcolor{black}{~\cite{Ado}
in the context of the AHE with massive Dirac fermions. As shown in
Sec.~\ref{sec:VII-CROSSING} for our model, the crossing diagrams
encoding quantum coherent skew scattering off two impurities provide
the dominant anomalous contribution over a wide range of parameters,
attaining remarkably large values away from the Gaussian regime.}

\section{MODEL SYSTEM\label{sec:III-MODEL SYSTEM}}

As a model system we consider a graphene sheet with extrinsic SOC
with origin in spin--orbit-active impurities. The low-energy physics
is captured by a Dirac Hamiltonian in two spatial dimensions with
a random impurity potential. It is convenient to introduce the SO(5)
representation of the spin algebra~\cite{Murakami,Zee} in terms
of $4\times4=1+5+10$ matrices, i.e., one identity, $\gamma^{0}$,
five $\gamma^{a}$ matrices, taken as $\gamma^{1}=\sigma_{1}\otimes s_{0}$,
$\gamma^{2}=\sigma_{2}\otimes s_{0}$, $\gamma^{3}=\sigma_{3}\otimes s_{3}$,
$\gamma^{4}=\sigma_{3}\otimes s_{2}$, and $\gamma^{5}=\sigma_{3}\otimes s_{1}$,
and ten adjoint matrices $\gamma^{ab}=i/2\,[\gamma^{a},\gamma^{b}]$,
where $\boldsymbol{\sigma}$ and $\boldsymbol{s}$ are Pauli matrices
defined in the sublattice and spin space, respectively. The Hamiltonian
density around the $K$ valley is given by 
\begin{equation}
\mathscr{H}=\psi^{\dagger}(\mathbf{x})\left\{ -i\,v\,\gamma^{j}\partial_{j}-\gamma_{0}\,\epsilon+V(\mathbf{x})\right\} \psi(\mathbf{x}),\label{eq:Ldirac}
\end{equation}
where $v$ is the Fermi velocity of charge carriers, $\epsilon$ is
the Fermi energy, and $V(\mathbf{x})$ denotes the disorder potential.
In this paper, we consider short-range impurity potentials of the
form 
\begin{equation}
V(\mathbf{x})=\sum_{i}M\,R^{2}\delta(\mathbf{x}-\mathbf{x}_{i})\,,\label{eq:impurity_pot}
\end{equation}
where $M$ is a $4\times4$ matrix encoding the spin and sublattice
texture of the impurity, and $R$ is a length scale mimicking a potential
range \cite{RS_Ferreira}. We note that, generally speaking, impurity
potentials in the continuum limit are described by enlarged $8\times8$
matrices accounting for the valley degree of freedom. In fact, when
the impurity range is of the order of the lattice spacing, intervalley
processes can counteract the intravalley skew scattering leading to
an overall reduction of the SH conductivity \cite{Pachoud}. In the
current work, we avoid additional complications arising from intervalley
scattering, and limit the discussion to the simplest model displaying
SHE. We therefore consider scattering potentials with ``intrinsic-type''
($\sigma_{3}s_{3}$) SOC \cite{Hernando06,Fabian09} 
\begin{equation}
M=\alpha_{0}\,\gamma_{0}+\alpha_{3}\,\gamma_{3}\,,\label{eq:M_matrix_intrinsic_SOC}
\end{equation}
where $\alpha_{3}$($\alpha_{0}$) is the SOC (electrostatic potential)
magnitude. \textcolor{black}{The intrinsic-type SOC conserves the
out-of-plane spin component, in addition to being an invariant of
the $C_{6v}$ point group, and thus it is the simplest form of SOC
in graphene; physical realizations include physisorbed atoms in the
hollow position, and randomly distributed top-position adatoms \cite{Pachoud}.
The presence of two different terms in the scattering potential is
responsible for a rich phenomenology, most noticeably a crossover
between a skew scattering- and a QSJ-dominated SHE in experimentally
accessible parameter regions, as demonstrated by the authors in Ref.~\cite{Milletari_1}.
In the following sections, we also show that the simultaneous presence
of two energy scales associated with the impurities leads to the breakdown
of commonly employed approximations. }

\textcolor{black}{Being interested in the effect of asymmetric and
strong scattering,} the standard \textit{\emph{Gaussian}} \textit{white
noise} approximation is not generally valid. In fact, and as we show
in Sec.~\ref{sec:IV-NON_CROSSING_WEAK_SCATT}, the Gaussian approximation
results into an $\epsilon$-independent anomalous contribution $\mathcal{Q}$.
To correctly take into account the role of the Fermi energy, we employ
the $T$-matrix approach introduced in Sec.~\ref{sec:II-METHODOLOGY}.
Within this approach, the self energy reads $\Sigma(\epsilon)=n\,\langle T(\epsilon)\rangle_{\textrm{dis}}$,
with the averaged $T$~matrix formally given by Eq.~\eqref{eq:Tmatrix_single}---its
diagrammatic representation is given in Fig.~\ref{fig:TMatrix}.
We find after some straightforward algebra 
\begin{equation}
\langle T(\epsilon)\rangle_{\textrm{dis}}=\frac{1}{2}\left(T_{+}+T_{-}\right)\gamma_{0}+\frac{1}{2}\left(T_{+}-T_{-}\right)\gamma_{3}\equiv T\,,\label{eq:TM_model}
\end{equation}
with 
\begin{equation}
T_{\pm}=\frac{R^{2}\,(\alpha_{0}\pm\alpha_{3})}{1-R^{2}\,(\alpha_{0}\pm\alpha_{3})\,g_{0}(\epsilon)}\equiv\epsilon_{\pm}\mp i\,\eta_{\pm}.\label{eq:TM1}
\end{equation}
In the above, $g_{0}(\epsilon)=-|\epsilon|/2\pi v^{2}\log\left(\Lambda/|\epsilon|\right)\mp i\,|\epsilon|/4v^{2}$
is the momentum integrated bare propagator in retarded (advanced)
sectors, and $\Lambda$ is a high energy cutoff \cite{RS_Ferreira}.
To simplify notation, in what follows we assume that the Fermi level
resides in the conduction band $\epsilon>0$. It is convenient to
decompose the self energy in real and imaginary part as 
\begin{align}
\Re\,\Sigma & =n(\delta\epsilon\,\gamma_{0}+m\,\gamma_{3})\,,\label{eq:Re_Sig}\\
-\Im\,\Sigma & =n(\eta\,\gamma_{0}+\bar{\eta}\,\gamma_{3})\,,\label{eq:Im_Sig}
\end{align}
with the following definitions: $\delta\epsilon=(\epsilon_{+}+\epsilon_{-})/2$,
$m=(\epsilon_{+}-\epsilon_{-})/2$, $\eta=(\eta_{+}+\eta_{-})/2$
and $\bar{\eta}=(\eta_{+}-\eta_{-})/2$. Here, $n\,\delta\epsilon$
is a chemical potential shift that can be reabsorbed in $\epsilon$,
while $n\,m$ is a (small) disorder-induced SOC gap. This result shows
that $\Sigma$ endows quasiparticles with two different lifetimes;
we have defined $n\,\eta$ and $n\,\bar{\eta}$ as the respective
energy and spin gap broadenings. The disorder averaged propagator
reads 
\begin{equation}
\mathcal{G}_{\mathbf{k}}^{R/A}(\epsilon)=\frac{(\epsilon\pm i\,n\,\eta)\gamma_{0}+n\,(m\mp i\,\bar{\eta})\gamma_{3}+v\,\gamma^{j}k_{j}}{(\epsilon\pm i\,n\,\eta)^{2}-n^{2}(m\mp i\,\bar{\eta})^{2}-v^{2}\,k^{2}}.\label{eq:avProp}
\end{equation}
In order to evaluate the SH conductivity, we also need the form of
the charge and spin current operators {[}Eqs.~\eqref{eq:charge_current}-\eqref{eq:spin_current}{]}.
In our model, these are given by $j_{y}^{z}=v/2\,\gamma_{13}$ and
$v_{x}=v\,\gamma_{1}$.

\section{GAUSSIAN DISORDER\label{sec:IV-NON_CROSSING_WEAK_SCATT}}

\subsection{Anomalous contribution\label{sub:IV-A-Anomalous_Gaussian}}

In this section we consider the weak scattering regime in the framework
of the so-called \textit{Gaussian approximation}. The aim is to show
the limitations of this widely used approximation. Consider Eq.~\eqref{eq:TM1}
for the $T$~matrix; expanding for $|\Re\,g_{0}\,R^{2}(\alpha_{0}\pm\alpha_{3})|\ll1$,
one obtains the first two diagrams of Fig.~\eqref{fig:TMatrix}.
Note that this is different from the naive expansion in the scattering
potentials $\alpha_{0}$ and $\alpha_{3}$, and allows us to treat
the two scattering mechanisms on equal footing. Keeping the second
(rainbow) diagram is equivalent to consider a random impurity potential
with a Gaussian white noise distribution~\cite{Rammer} 
\begin{align}
\langle V(\mathbf{x})\rangle_{\textrm{dis}} & =0\,,\\
\langle V(\mathbf{x})V(\mathbf{x}')\rangle_{\textrm{dis}} & =n\,R^{4}\,M^{2}\,\delta(\mathbf{x}-\mathbf{x}').\label{eq:gaussiandis}
\end{align}
This model has been widely used to study disordered systems. Note
that the zero average condition of the random potential comes from
the fact that the first diagram in the $T$~matrix expansion involves
only the real part of the self energy. In standard (parabolic) systems,
one can generally re-adsorb the real part of the self energy in a
redefinition of the Fermi energy; therefore, one can always recenter
the distribution around zero. In the present case, the real part of
the self energy also contains a random spin gap term (absent in the
clean Hamiltonian) and therefore it cannot be renormalized away. However,
at the level of the Gaussian approximation, one can show that adding
the random spin gap term does not modify the leading order result
for the SH conductivity, therefore we will ignore this term henceforth.
The imaginary part of the self energy is given by the Born limit expression
\begin{align}
\Im\Sigma(\epsilon) & =\langle V(\mathbf{x})V(\mathbf{x}')\rangle_{d}\,\Im\,G_{0}(\mathbf{x},\mathbf{x}';\epsilon)\label{eq:Im_S}\\
 & =n\,R^{4}\,M^{2}\frac{\epsilon}{4v^{2}}=n(\eta\,\gamma_{0}+\bar{\eta}\,\gamma_{3}),\label{eq:Im_S_exp}
\end{align}
that reproduces the definition for the imaginary part of the self
energy given in Sec.~\ref{sec:III-MODEL SYSTEM}. However, in the
Gaussian model the actual values of the energy broadening parameters
are $\eta_{\pm}\simeq R^{4}(\alpha_{0}\pm\alpha_{3})^{2}\epsilon/(4v^{2})$.
\begin{figure}[t!]
\centering{}\includegraphics[clip,width=0.85\columnwidth]{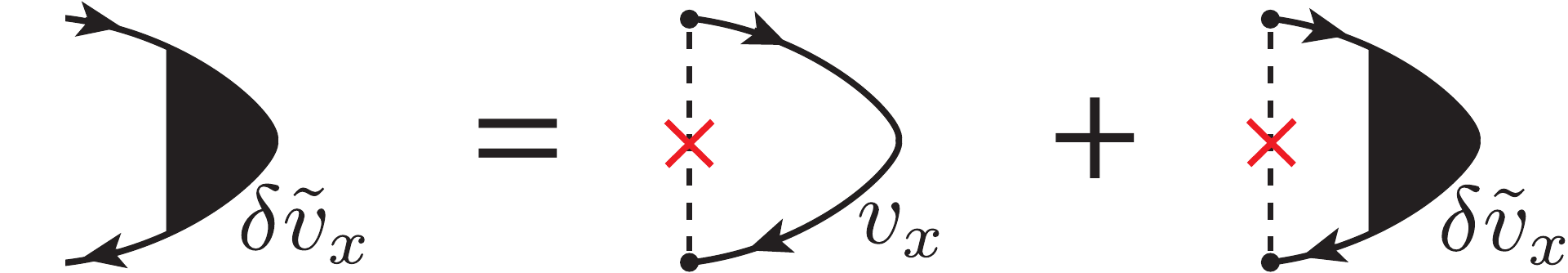}
\caption{Bethe-Salpeter equation for the standard vertex correction $\delta\tilde{v}_{x}$.
Black dots connected by a dashed line represent impurity potential
($\hat{M}$) insertions, while the red $\times$ represent an impurity
density insertion.}
\label{fig:vertexc} 
\end{figure}

In order to study the transport properties of the system, one needs
to consider disorder insertions in the $4$-point function, leading
to the \textit{\emph{vertex corrections}}. At the level of the Gaussian
approximation, impurity scattering contributes to a single ladder
diagram (Fig.~\ref{fig:vertexc}), connecting the advanced and retarded
sector of the response function. The renormalized vertex can be expressed
as $\tilde{v}_{x}=v_{x}+\delta v_{x}$, where $v_{x}$ is the bare
vertex and $\delta v_{x}$ are the corrections due to impurity scattering.
In order to take into account multiple independent scatterings, one
needs to resum the ladder series self-consistently\textcolor{black}{{}
by writing the Bethe-Salpeter equation} 
\begin{equation}
\delta v_{x}=\bar{v}_{x}+n\,R^{4}\sum_{\mathbf{k}}M\,\mathcal{G}_{\mathbf{k}}^{R}\,\delta v_{x}\,\mathcal{G}_{\mathbf{k}}^{A}\,M\,,\label{eq:delta_v_x}
\end{equation}
where $\bar{v}_{x}$ contains only the ladder part, see Fig.~\ref{fig:vertexc}.
At leading order in the impurity density we find 
\begin{align}
\bar{v}_{x} & =n\,R^{4}\,\int\frac{d^{2}\mathbf{k}}{(2\pi)^{2}}\left\{ M\,\mathcal{G}_{\mathbf{k}}^{R}\,v_{x}\,\mathcal{G}_{\mathbf{k}}^{A}\,M\right\} \label{eq:bar_v_x}\\
 & =v\,(a\,\gamma_{1}+b\,\gamma_{13})\,,\label{eq:bar_vx_2}
\end{align}
with

\begin{equation}
a\simeq\frac{(\alpha_{0}^{2}-\alpha_{3}^{2})}{2(\alpha_{0}^{2}+\alpha_{3}^{2})}\,,\quad b\simeq\frac{n\,R^{4}}{2v^{2}}\frac{\alpha_{0}\,\alpha_{3}(\alpha_{0}^{2}-\alpha_{3}^{2})}{(\alpha_{0}^{2}+\alpha_{3}^{2})}.\label{eq:ladder}
\end{equation}
Note that the $b$ coefficient starts at order $n$, while $a$ is
independent of $n$. The only matrix elements contributing to the
vertex renormalization are those proportional to $\gamma_{1}$ and
$\gamma_{13}$. This means that we can decompose the vertex part in
the second diagram of Fig.~\ref{fig:vertexc} as $\delta v_{x}=\delta v_{x}^{1}\,\gamma_{1}+\delta v_{x}^{2}\,\gamma_{13}$.
Using this ansatz in the Bethe-Salpeter equation for the vertex part,
and taking the trace of $\delta v_{x}$ together with $\gamma_{1}$
or $\gamma_{13}$, we obtain 
\begin{equation}
\delta v_{1}=\frac{v\left(\alpha_{0}^{2}-\alpha_{3}^{2}\right)}{\alpha_{0}^{2}+3\,\alpha_{3}^{2}},\,\delta v_{2}=2\,n\,R^{4}\frac{\alpha_{0}\,\alpha_{3}\left(\alpha_{0}^{4}-\alpha_{3}^{4}\right)}{v\left(\alpha_{0}^{2}+3\,\alpha_{3}^{2}\right)^{2}}.\label{eq:vertices}
\end{equation}
In this way, the renormalized vertex can be written as $\tilde{v}_{x}=(v+\delta v_{1})\,\gamma_{1}+\delta v_{2}\,\gamma_{13}$.
Using the renormalized vertex into the expression for the SH-conductivity,
and multiplying by a factor of $2$ valley degeneracy, we finally
obtain the SH conductivity in the noncrossing approximation 
\begin{align}
\left.\sigma_{\textrm{SH}}^{\textrm{nc}}\right| & _{\textrm{Gauss.}}=2\int\frac{d^{2}\mathbf{k}}{(2\pi)^{2}}\:\textrm{Tr}\left[\,j_{y}^{z}\,\mathcal{G}_{\mathbf{k}}^{R}(\epsilon)\,\tilde{v}_{x}\,\mathcal{G}_{\mathbf{k}}^{A}(\epsilon)\,\right]\label{eq:SHdressed}\\
 & \qquad\;=\frac{8e^{2}}{h}\,\frac{\alpha_{0}\,\alpha_{3}\,(\alpha_{0}^{2}+\alpha_{3}^{2})}{(\alpha_{0}^{2}+3\,\alpha_{3}^{2})^{2}}\equiv Q_{\textrm{nc}}^{\textrm{G}}\,.\label{eq:SH_Gauss_Q}
\end{align}
This result suggests an energy independent SH conductivity. However,
as we discuss in Ref.~\cite{Milletari_1}, this is an artifact of
the Gaussian model. The limitations of the Gaussian approximation
and their implications for a correct analysis of the extrinsic SHE
will be discussed in Sec.~\ref{sub:VI-A-NonCroossing_Full_Diagramm}.
We finalize this section by pointing out the striking similarity between
the above result and the noncrossing Hall conductivity for the 2D
massive Dirac band with Gaussian scalar disorder~\cite{Synitsin_link,Ado}.
The expression for the Hall conductivity $\sigma_{xy}$---see e.g.,
Eq.~(3) in Ref.~\cite{Ado}---can be recovered multiplying the right-hand
side of Eq.~\eqref{eq:SH_Gauss_Q} by $(-1)\times1/2$ (a minus sign
to obtain a $xy$ response, and a $1/2$ factor to remove the valley
degeneracy), and sending $\alpha_{0}\rightarrow\epsilon$ and $\alpha_{3}\rightarrow m$,
where $m$ in the AHE represents the band gap. The straightforward
mapping between the two results is not accidental and can be traced
back to the similar structure of the dressed propagators in both models,
see Eq.~\eqref{eq:avProp}.

\subsection{Skew Scattering \label{sub:IV-B-Skew_Weak}}

\begin{figure}[!t]
\centering{}\includegraphics[width=0.85\columnwidth]{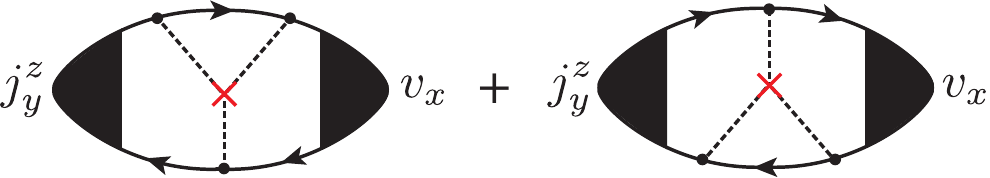} \caption{\label{fig:Ybubb}``$Y$'' diagrams contributing at order\textbf{
$V^{3}$}. The dashed lines represent contractions of the impurity
potentials and the red $\times$ represent an impurity density insertion.}
\end{figure}

Within the weak Gaussian approximation it is not possible to assess
the effect of skew scattering events. This is easy to understand by
expanding the $T$~matrix: \textcolor{magenta}{{} }\textcolor{black}{the
Gaussian term can only describe the width of the disorder distribution
but not its skewness. }For this reason, one needs to include order
$V^{3}$ terms {[}see also Sec.\,\ref{sub:VI-B-Boltzmann}{]}. In
the standard treatments, this is done by allowing for a ``non-Gaussian''
average of the form 
\begin{equation}
\langle V(\mathbf{x})V(\mathbf{x}')V(\mathbf{x}'')\rangle_{\textrm{dis}}=n\,R^{6}\,M^{3}\,\delta(\mathbf{x}-\mathbf{x}')\,\delta(\mathbf{x}'-\mathbf{x}'').\label{eq:nongauss}
\end{equation}
It should be noted that in the standard approach, three-point correlators
are only used to evaluate the disorder average of the 4-point function
but not of the self energy, that is still evaluated at the Gaussian
level. Here we follow this approach and show that it leads indeed
to a consistent result in the weak scattering regime. Using the ``non-Gaussian
`` average prescription of Eq.~\eqref{eq:nongauss} in the four
point function, one obtains the so called $Y$~diagrams represented
in Fig.~\ref{fig:Ybubb} {[}$b_{2}$ and $b_{3}$ in Fig.~\ref{fig:general_exp}{]}.
In the spirit of the perturbative approach of this section, ``$Y$''
insertions and vertex corrections are treated as if they represent
two separate processes. On the other hand, in Ref.~\cite{Milletari_1}
we showed that ``$Y$'' insertions are themselves part of the vertex
corrections. We will return to this important issue in the following
section. Finally, note that in the perturbative approach one also
needs to dress the spin vertex $\tilde{j}_{y}^{z}=(v+\delta v_{1})/2\,\gamma_{13}+\delta v_{2}/2\,\gamma_{1}$
(see Fig.~\ref{fig:Ybubb}).

It is convenient to recast the expression for the conductivity in
terms of proper spin ($\Gamma_{y}^{z}$) and charge ($\Gamma_{x}$)
vertices 
\begin{align}
\sigma_{Y} & =n\,R^{6}\:\textrm{Tr}\left\{ \Gamma_{y}^{z}\,g^{R}\,M\,\Gamma_{x}+\Gamma_{y}^{z}\,\Gamma_{x}\,M\,g^{A}\right\} \,,\label{eq:Ybub1}
\end{align}
where 
\begin{align}
\Gamma_{y}^{z} & =\int\frac{d^{2}\mathbf{k}}{(2\pi)^{2}}\,M\,\mathcal{G}_{\mathbf{k}}^{A}\,\tilde{j}_{y}^{z}\,\mathcal{G}_{\mathbf{k}}^{R}\,M\,,\label{eq:Gamma_y_z}\\
\Gamma_{x} & =\int\frac{d^{2}\mathbf{k}}{(2\pi)^{2}}\,\mathcal{G}_{\mathbf{k}}^{R}\,\tilde{v}_{x}\,\mathcal{G}_{\mathbf{k}}^{A}\,,\label{eq:Gamma_x}
\end{align}
where $g^{R/A}$ are the integrated, dressed Green functions 
\begin{align}
g^{R/A} & \simeq-\frac{(\epsilon\pm\imath\,n\,\eta)\gamma_{0}+n(m\mp\imath\,\bar{\eta})\gamma_{3}}{4\pi\,v^{2}}\times\nonumber \\
 & \qquad\left[\pm\imath\,\pi+2\,\log\left(\frac{\Lambda}{\epsilon}\right)\right]\,.\label{eq:dressed_int_prop}
\end{align}
At leading order we obtain 
\begin{align}
\Gamma_{y}^{z} & \simeq\frac{R^{4}\,\alpha_{0}\,\alpha_{3}\left(\alpha_{0}^{4}-\alpha_{3}^{4}\right)}{v\left(\alpha_{0}^{2}+3\,\alpha_{3}^{2}\right)^{2}}\gamma_{1}+\frac{v(\alpha_{0}^{2}-\alpha_{3}^{2})}{2\,n\,\left(\alpha_{0}^{2}+3\,\alpha_{3}^{2}\right)}\gamma_{13}\,,\label{eq:vertY}\\
\Gamma_{x} & \simeq\frac{v}{n\,R^{4}\left(\alpha_{0}^{2}+3\,\alpha_{3}^{2}\right)}\gamma_{1}+\frac{2\,\alpha_{0}\,\alpha_{3}\left(\alpha_{0}^{2}+\alpha_{3}^{2}\right)}{v\left(\alpha_{0}^{2}+3\,\alpha_{3}^{2}\right)^{2}}\gamma_{13}\,.\label{eq:verteX}
\end{align}
These are the terms responsible for the semiclassical skew scattering
term in Eq.~\eqref{eq:SH_imp_dens_expansion}. However, we would
like to remark that order $\mathcal{O}(n^{0})$ terms resulting from
Eq.~(\ref{eq:Ybub1}) should in principle be added to the anomalous
contribution coming from the empty bubble. At leading order in the
impurity density we find 
\begin{equation}
\sigma_{Y}=\frac{2e^{2}}{h}\,\frac{\epsilon\,\alpha_{3}(\alpha_{0}^{2}-\alpha_{3}^{2})}{n\,R^{2}\,(\alpha_{0}^{2}+3\,\alpha_{3}^{2})^{2}}.\label{eq:yweak}
\end{equation}
In the perturbative approach, the skew scattering contribution to
$\sigma_{\textrm{SH}}$ increases linearly with the energy. In addition,
this result predicts that the skew scattering vanishes identically
for potentials satisfying $\alpha_{0}=\pm\alpha_{3}$. As shown in
the next section, this is an exact symmetry of the model that is preserved
at all orders in perturbation theory.

\section{BEYOND GAUSSIAN Approximation\label{sec:V-NONPERTURBATIVE}}

\subsection{Nonperturbative diagrammatic approach \label{sub:VI-A-NonCroossing_Full_Diagramm}}

In this section we review the self-consistent treatment of SH response
functions introduced by the authors in Ref.~\cite{Milletari_1}.
The self-consistent calculation of $\sigma_{\textrm{SH}}$ cures the
spurious energy-independent anomalous contribution obtained in the
Gaussian approximation, and also provides an expression valid for
arbitrarily strong scattering potentials. In this approach, one uses
the full $T$~matrix both in the self energy and in the 4-point function.
The $T$~matrix provides a resummation of all moments of the random
disorder distribution in the Markovian (uncorrelated) approximation~\cite{Rammer,Rammer2}.
In general, the distribution of a random variable can be defined by
its moments: the ``Gaussian'' term (or variance) only contains informations
about the width of the distribution (the deviation from the average),
while the third (skewness) and fourth (kurtosis) moments, give informations
about its actual shape~\cite{Atland}. Higher order moments further
define the shape of the distribution. In Section~\ref{sub:IV-A-Anomalous_Gaussian},
we have used the Born criterion to perform a moment expansion up to
second order. In its common (and widely used) form, the Born criterion
is a statement about the \emph{magnitude} of the scattering potential
and its validity is justified on purely perturbative grounds. This
simple criterion has been used to justify the evaluation of the anomalous
term (QSJ) by keeping just the second moment. While this argument
holds for Hamiltonians with trivial symmetry structure or a single
energy scale, it is not generally true otherwise. In our model, for
example, by keeping only the Gaussian moment in the $T$~matrix expansion,
the resulting spin-current only accounts for symmetric scattering
processes. Below, we show that by keeping higher order moments of
the disorder distribution, one can access physically distinguishable
processes contributing to the SH current at the same order in $\alpha_{0}$
and $\alpha_{3}$ as the Gaussian one (QSJ).

As shown in Fig.~\ref{fig:general_exp}, the expansion of the $T$~matrix
in the 4-point function {[}second line of Eq.~\eqref{eq:sH_cond_Baym}{]},
corresponds to a series of two-particle noncrossing diagrams containing
multiple insertions of the bare scattering potential. The ladder corrections
yielding the QSJ contribution and the $Y$ diagrams describing skew
scattering in the weak scattering regime have been already considered
in Sec.~\ref{sec:IV-NON_CROSSING_WEAK_SCATT}. In order to describe
skew scattering and anomalous processes on equal footing, we need
to solve for the complete 4-point function. Diagrammatically, the
latter corresponds to the full series of topologically equivalent
two-particle noncrossing diagrams shown in the first two lines of
Fig.~\ref{fig:general_exp}. As explained in Sec.~\ref{sec:II-METHODOLOGY},
this is done by introducing a fully dressed vertex function by exchanging
$M\to T^{\lambda}$ in Eq.~\eqref{eq:ladder}. We find 
\begin{figure}[t!]
\centering{}\includegraphics[clip,width=0.85\columnwidth]{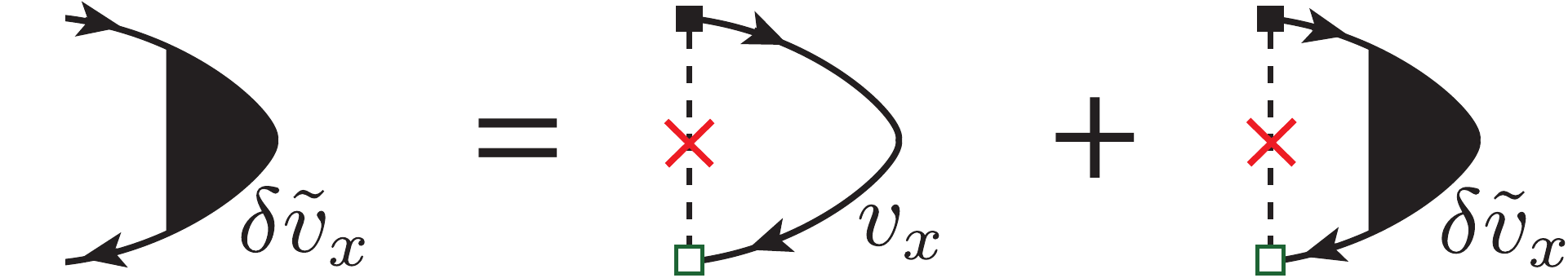}
\caption{Bethe-Salpeter equation for the full vertex correction $\delta\tilde{v}_{x}$.
Full (open) squares connected by a dashed line represent $T$ ($T^{*}$)
matrix insertions, while the red $\times$ represent an impurity density
insertion.}
\label{fig:vertexc-full} 
\end{figure}

\begin{align}
\bar{v}_{x} & =n\,\int\frac{d^{2}\mathbf{k}}{(2\pi)^{2}}\:\hat{T}\,\mathcal{G}_{\mathbf{k}}^{R}\,v_{x}\,\mathcal{G}_{\mathbf{k}}^{A}\,\hat{T}^{*}=v\,(a\,\gamma_{1}+b\,\gamma_{13})\,,\label{eq:Tladder}
\end{align}
and 
\begin{align}
a & \simeq\epsilon\,\frac{\eta_{+}\,\eta_{-}+\epsilon_{+}\,\epsilon_{-}}{4v^{2}(\eta_{+}+\eta_{-})}-n\,f_{a}(\eta_{+},\eta_{-},\epsilon_{+},\epsilon_{-}),\label{eq:a}\\
b & \simeq\epsilon\,\frac{\eta_{+}\,\epsilon_{-}-\eta_{-}\,\epsilon_{+}}{4v^{2}(\eta_{+}+\eta_{-})}+n\,f_{b}(\eta_{+},\eta_{-},\epsilon_{+},\epsilon_{-}),\label{eq:b}
\end{align}
where $f_{a}$ and $f_{b}$ are complicated functions of $\eta_{\pm},\epsilon_{\pm}$;
explicit expressions are given in Appendix~\ref{a:SELFCONS}.

Comparing with the Gaussian expression of Eq.~\eqref{eq:ladder},
we see a qualitative difference, namely that the coefficient $b$
now contains a term that is independent of the impurity density $n$.
It is evident that the self consistent method treats on equal footing
the charge and the spin vertex (note the similar structure of $a$
and $b$ coefficients). Indeed, it is easy to see that vertex corrections
generate an effective spin-spin ($j_{y}^{z}-j_{y}^{z}$) current response
function that is ultimately responsible for skew scattering, $\mathcal{S}(\epsilon)$.
We anticipate here that this additional term in $b$ is also responsible
for the non vanishing of the crossing $\Psi$-diagrams evaluated in
Sec.\,\ref{sec:VII-CROSSING}.

Solving the Bethe-Salpeter equation for the\textit{ $T$~}\textit{\emph{matrix-dressed}}\textit{
}\textit{\emph{vertex }}\textcolor{black}{(see Fig.~\ref{fig:vertexc-full})
}we see that the additional term in $b$ responsible for skew scattering
also influences $\mathcal{Q}$, meaning that skew scattering and QSJ
mechanisms are never truly separated. In Ref.~\cite{Milletari_1}
we obtained the full SH conductivity in the noncrossing approximation
\begin{align}
\sigma_{\textrm{SH}} & =\frac{\epsilon\,\delta v_{20}}{2\,n\,v\,\eta}+\Big\{\frac{\epsilon\,\delta v_{22}+2\,(v+\delta v_{10})\,\bar{\eta}}{2\,v\,\eta}\nonumber \\
 & -\delta v_{20}\left(\frac{1}{\pi v}+\frac{\bar{\eta}\,m}{2\,v\,\eta^{2}}\right)\Big\}\equiv\mathcal{S}(\epsilon)/n+\mathcal{Q}_{\textrm{nc}}(\epsilon),\label{eq:FullSH}
\end{align}
in units of $e^{2}/h$. The explicit form of the vertex corrections
$\delta v_{ij}$ is given in Appendix~\ref{a:SELFCONS} (for calculation
details refer to the supplemental material in Ref.~\cite{Milletari_1}).
\begin{figure}[t!]
\centering{}\includegraphics[width=1\columnwidth]{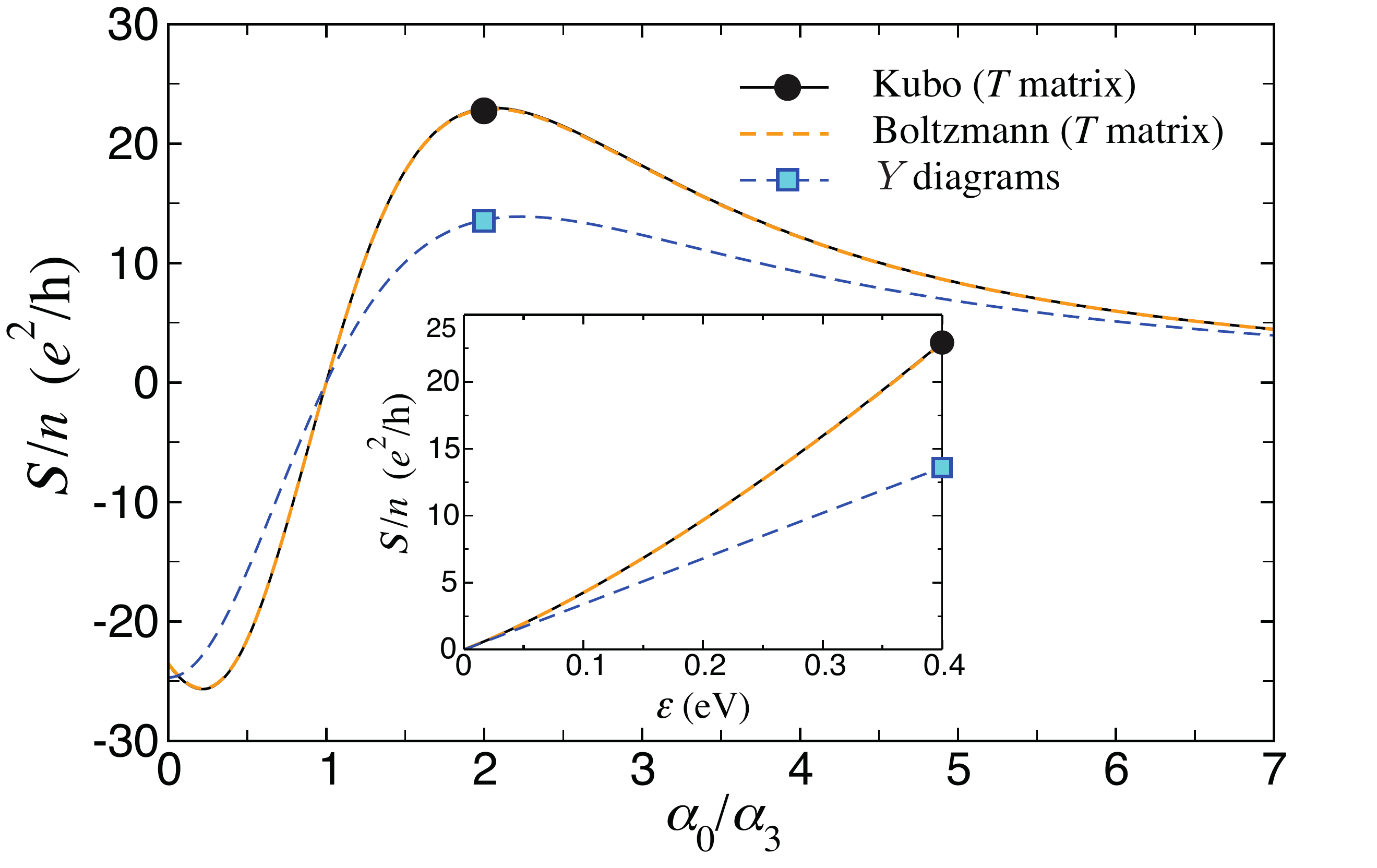} \caption{\label{fig:skew}Semiclassical skew scattering contribution, $\mathcal{S}(\epsilon)/n$.
The various approaches: self-consistent LRT, ``$Y$''-diagrams approximation,
and Boltzmann prediction. The calculation has $R=6$~nm, $\alpha_{3}=10$~meV,
$n=10^{12}$~cm$^{-2}$, and $\epsilon=0.4$~eV. The inset shows
the variation with the Fermi energy for $\alpha_{0}=20$~meV.}
\end{figure}

\subsubsection*{Discussions}

We begin our discussions with the skew scattering contribution to
the SH conductivity, $S(\epsilon)/n$. The improvement over the weak
scattering (``$Y$''-diagrams) approximation is borne out in Fig.~\ref{fig:skew},
which shows the SH conductivity generated by impurities with a large
radius (in our model this is tantamount to a strong scattering potential).
One sees that ``$Y$'' diagrams fail to provide an accurate result,
even though the basic trends are still captured. (We expect even stronger
discrepancies to arise in models displaying multiple impurity resonances,
such as the clustered-adatom/nanoparticle potentials considered in
Ref.~\cite{Ferreira14}.) We finally note that for our delta-impurity
model, the range of validity of the ``$Y$''-diagrams approximation
is linked to the strength of the SOC and reads $|\Re\,g_{0}R^{2}\alpha_{3}|\ll1$.
When $\alpha_{0}=\pm\alpha_{3}$ the vertex function $\delta v_{20}$
vanishes identically and thus $\mathcal{S}(\epsilon)=0$. \textcolor{black}{This
situation corresponds to a potential localized on one of the sublattices,
which always leads to symmetric scattering amplitudes; see Sec.~\ref{sub:VI-B-Boltzmann}.}

We now discuss the anomalous term, historically associated to QSJ
events. Two approaches are normally employed: (i) the standard diagrammatic
approach with the empty bubble dressed with ladder diagrams (Sec.~\ref{sub:IV-A-Anomalous_Gaussian}),
and (ii) the generalized semiclassical transport equations accounting
for corrections to the distribution function and velocity operator
arising from side-jump accumulation~\cite{Synitsin_coordinate_shift}.
However, the QSJ is far from being the only mechanism yielding an
anomalous contribution. For instance, as noted by Synitsin~\cite{Synitsin_review},
incoherent multiple skew scattering resulting from the dressing of
the wavefunctions with an average SOC also leads to a parametrically
equivalent contribution. In the Boltzmann formalism, the latter can
be incorporated heuristically by means of a virtual crystal approximation,
$H_{0}\rightarrow H_{0}+\Sigma$. The transition rates are then computed
with respect to self-energy dressed wavefunctions, which results in
anomalous-type $O(n^{0})$ corrections. Within the quantum LRT formalism,
the self energy correction is naturally accounted for in the disordered
Green functions. Since the average SOC is linear in the impurity density,
incoherent multiple scattering can usually be safely neglected. In
the full noncrossing result {[}Eq.~\eqref{eq:FullSH}{]} this is
the term proportional to the spin gap, $m$. There are other mechanisms
that can lead to a meaningful contribution, though, of the same order
of QSJ or even stronger. As mentioned in the Introduction, quantum
coherent multiple scattering (encoded in two-particle crossing diagrams)
gives rise to an anomalous contribution. This has been recently discovered
in a minimal model of the AHE with Gaussian disorder~\cite{Ado}.
We will confirm the crucial role played by quantum coherent processes
in Sec.~\ref{sec:VII-CROSSING}.

A careful inspection of $\mathcal{Q}_{\textrm{nc}}$ discloses an
anomalous contribution that cannot be linked to any of the aforementioned
processes---this is one of the central results of this work. In order
to establish the origin of this new contribution, it is convenient
to compare the full noncrossing expression with the Gaussian result,
$\mathcal{Q}_{\textrm{nc}}^{\textrm{G}}$ {[}Eq.~\eqref{eq:SH_Gauss_Q}{]}.
In addition to a weak Fermi energy dependence absent in the Gaussian
approximation, $\mathcal{Q}_{\textrm{nc}}$ contains qualitatively
different terms. These terms arise from different physical processes
of the same order of the ``Gaussian'' ones and therefore cannot
be neglected a priori. In order to see this explicitly, we consider
the weak scattering limits of $\mathcal{Q}_{\textrm{nc}}$ in the
formal $\epsilon\to0$ limit 
\begin{equation}
\mathcal{Q}_{\textrm{nc}}(\epsilon\to0)\simeq\left\{ \begin{array}{cc}
8\,(\alpha_{3}/\alpha_{0}) & \,,\:\alpha_{3}\ll\alpha_{0}\\
\frac{40}{27}\,(\alpha_{0}/\alpha_{3}) & \,,\:\alpha_{3}\gg\alpha_{0}
\end{array}\,,\right.\label{eq:limit2}
\end{equation}
The above expression deserves a comment. Although the $\epsilon\to0$
limit can be formally taken in Eq.~\eqref{eq:FullSH}, in this limit
$k_{F}l\rightarrow0$ and the original perturbative series diverges,
i.e., terms with higher powers of $n$ become increasingly important.
Ignoring this technicality, we find that the Gaussian result {[}Eq.~\eqref{eq:SH_Gauss_Q}{]}
is only recovered for $\alpha_{3}\ll\alpha_{0}$, whereas in the opposite
limit, $\alpha_{3}\gg\alpha_{0}$, the Gaussian approximation yields
$Q_{\textrm{nc}}^{\textrm{G}}\simeq8/9\,(\alpha_{0}/\alpha_{3})$,
which differs from the result in Eq.~\eqref{eq:limit2}. The former
limit corresponds to the case when the SOC term is smaller than any
other energy scale in the system. In this case, the parity symmetry
breaking moments of the disorder distribution give a negligible contribution
and higher moments can be safely neglected. This limit corresponds
to the standard Born regime. In the opposite limit, the SOC term becomes
dominant, meaning that parity symmetry breaking moments \emph{must}
be included. The resummation scheme used to obtain Eq.~\eqref{eq:FullSH}
captures this important nonperturbative feature. Finally, we would
like to stress once more that, since the additional processes contributing
to Eq.~\eqref{eq:FullSH} are not included in the Gaussian approximation,
it is not possible to recover the Gaussian result in the $\alpha_{3}\gg\alpha_{0}$
limit. We have checked that higher order potential insertions in the
vertex corrections are responsible for the different pre-factor in
Eq.~\eqref{eq:limit2}. We thus attribute the enhanced anomalous
contribution reported here to skew scattering corrections to QSJ processes.

\subsection{Semiclassical approach: skew scattering\label{sub:VI-B-Boltzmann}}

Semiclassical transport theory provides a simple framework to tackle
electronic transport in materials. Broadly speaking, semiclassical
approaches are expected to be accurate in the $k_{F}l\gg1$ limit,
where the use of classical distribution functions $f(\mathbf{x},\mathbf{p},t)$
is justified \cite{Haug_Jauho_QuantumKinetics}. The link between
Boltzmann transport theory and the quantum diagrammatic approach has
been illustrated recently for 2D massive Dirac fermions in the context
of the AHE \cite{Synitsin_link}. In that work, a generalization of
the standard Boltzmann transport equations (BTEs) informed by an elegant
adiabatic semiclassical wavepacket dynamics analysis \cite{Synitsin_coordinate_shift,Synitsin_review}
is employed to assess the skew scattering and side-jump contributions
to the conductivity. The results obtained from the generalized BTEs
are then matched one-to-one to particular Kubo--Streda diagrams. It
is important to note that the correspondence between the two formalisms
in Ref.~\cite{Synitsin_link} is established for a simple (scalar)
impurity model and limited to the weak scattering regime. Our findings
in the previous section clearly show that weak Gaussian approximations
break down due to the intricate correlated nature of QSJ and skew
scattering arising from the nontrivial structure of the impurity potential.
This suggests that a simple correspondence between the anomalous contribution
$\mathcal{Q}(\epsilon)$ obtained from generalized BTEs and the rigorous
quantum diagrammatic technique may not exist in general. Nevertheless,
one can use Boltzmann theory to evaluate the leading (semiclassical)
term in the conductivity expansion Eq.~(\ref{eq:SH_imp_dens_expansion}).

\begin{figure}[t]
\includegraphics[width=0.7\columnwidth]{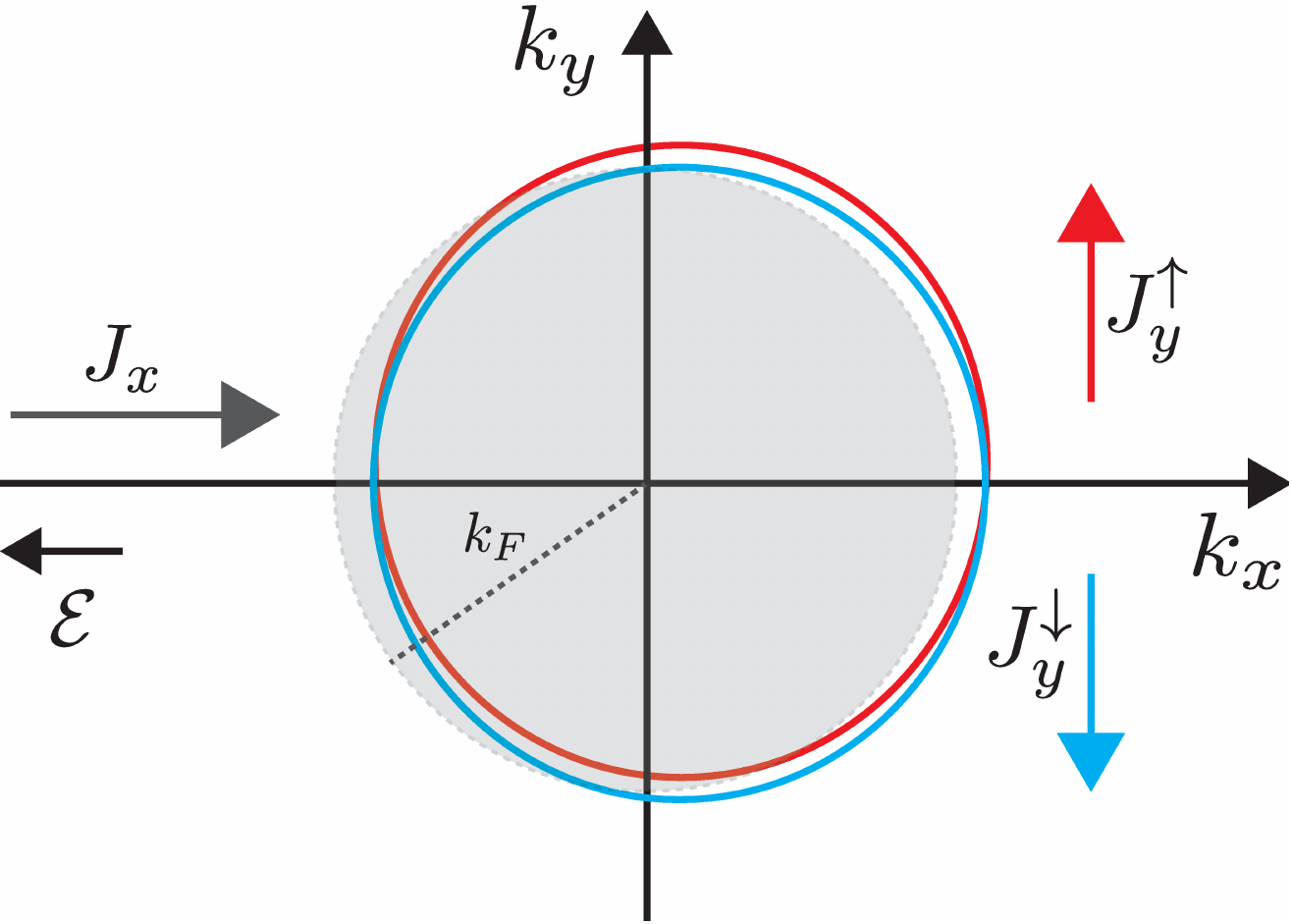} \caption{\label{fig:Fermi_Surf_Distortion}Schematic representation of the
distortion of the Fermi surface due to an external electric field
in the presence SOC, and the resulting build up of longitudinal charge
and transverse spin currents.}
\end{figure}

In what follows, we show by explicit calculation that the skew scattering
contribution computed by means of an exact solution of linearized
BTEs \emph{quantitatively agrees} with the non-perturbative diagrammatic
calculation over a wide range of scattering regimes. The starting
point of our semiclassical analysis is the standard BTE for the electronic
motion in an uniform system 
\begin{equation}
\frac{\partial n_{\sigma}}{\partial t}+\dot{\mathbf{k}}\cdot\partial_{\mathbf{k}}n_{\sigma}=\;\mathcal{I}[n_{\sigma}]\,.\label{eq:BTEs}
\end{equation}
In the above, $n_{\sigma}\equiv n_{\sigma}(\mathbf{k},t)$ is the
distribution function, $\sigma=\pm1$ {[}$\uparrow$($\downarrow$){]}
labels the spin projection, $\dot{\mathbf{k}}\equiv d\mathbf{k}/dt$,
and $\mathcal{I}[.]$ denotes the collision integral. For a small
external perturbation, the linearized BTEs characterizing the steady
state for up and down spin species read as 
\begin{equation}
-\mathcal{E}v_{\mathbf{k},x}\left(\frac{\partial n^{0}}{\partial\epsilon}\right)_{\epsilon=\epsilon(\mathbf{k})}=\;\mathcal{I}[n_{\sigma}(\mathbf{k})]\,,\label{eq:linearized_BTEs}
\end{equation}
where we used the classical equation of motion $\dot{\mathbf{k}}=-\mathcal{E}\hat{x}$
to simplify the expression, $\mathbf{v}_{\mathbf{k}}$ denotes the
band velocity of pristine graphene, \emph{i.e.}, $v_{\mathbf{k}}=v(\cos\theta_{\mathbf{k}},\sin\theta_{\mathbf{k}})^{\textrm{t}}$,
and $n^{0}$ is the equilibrium Fermi-Dirac distribution function.
In our model with spin-conserving impurities, the collision integral
does not mix opposite spins, and hence it reduces to its familiar
form in paramagnetic systems 
\begin{equation}
\mathcal{I}[n_{\sigma}(\mathbf{k})]=\sum_{\:\mathbf{k}^{\prime}}\left[n_{\sigma}(\mathbf{k}^{\prime})-n_{\sigma}(\mathbf{k})\right]W_{\sigma}(\mathbf{k},\mathbf{k}^{\prime})\,,\label{eq:collision_integral}
\end{equation}
with quantum-mechanical transition probability given by the generalized
Fermi golden rule 
\begin{equation}
W_{\sigma}(\mathbf{k}^{\prime},\mathbf{k})=2\pi n\,|T_{\mathbf{k}^{\prime}\mathbf{k}}^{\sigma}|^{2}\:\delta(\epsilon_{\mathbf{k}}-\epsilon_{\mathbf{k}^{\prime}})\,,\label{eq:Fermi_Golden_Rule}
\end{equation}
where $T_{\mathbf{k}^{\prime}\mathbf{k}}^{\sigma}=\langle u_{\mathbf{k}^{\prime}},\sigma|\hat{T}|u_{\mathbf{k}},\sigma\rangle$,
and $|u_{\mathbf{k}},\sigma\rangle=e^{-i\mathbf{k}\cdot\mathbf{r}}|\psi_{\mathbf{k}}^{0}\rangle\otimes|\sigma\rangle$
denotes conduction-band spinors. Using the explicit form of the $T$~matrix
for our model {[}see Eq.~(\ref{eq:TM1}){]} and $|u_{\mathbf{k}}\rangle=(1,e^{i\theta_{\mathbf{k}}})^{\textrm{t}}$,
we easily find 
\begin{equation}
T_{\mathbf{k}^{\prime}\mathbf{k}}^{\sigma}=\sum_{p=\pm1}\frac{T_{p}}{4}\left[\left(1+p\sigma\right)+\left(1-p\sigma\right)e^{-i\theta}\right]\equiv T^{\sigma}(\theta)\,,\label{eq:T_matrix_eigen}
\end{equation}
where $\theta=\theta_{\mathbf{k}^{\prime}}-\theta_{\mathbf{k}}$ is
the scattering angle, and $T_{\pm}$ as given in Eq.~(\ref{eq:TM1}).
From the above expression, it is clear that the scattering amplitude
probabilities possess a right--left asymmetry $|T^{\sigma}(\theta)|\neq|T^{\sigma}(-\theta)|$
as long as $T_{+}\neq T_{-}$, which occurs whenever $\alpha_{3}\neq0$
and $\alpha_{3}\neq\alpha_{0}$. The solution of Eq.~(\ref{eq:linearized_BTEs})
is given by $n_{\sigma}(\mathbf{k})=n^{0}+\delta n_{\sigma}(\mathbf{k})$,
where $\delta n_{\sigma}(\mathbf{k})$ consists of two parts 
\begin{equation}
\delta n_{\sigma}(\mathbf{k})=A_{\parallel}\cos\theta_{\mathbf{k}}+A_{\perp}\sin\theta_{\mathbf{k}}\,,\label{eq:sol_BTE}
\end{equation}
a standard longitudinal transport term $A_{\parallel}\propto\tau_{xx}$,
and a skew scattering contribution, $A_{\perp}\propto\tau_{yx}$.
The establishment of a net transverse spin current due to the antisymmetric
term in the distribution function Eq.~(\ref{eq:sol_BTE}) is illustrated
in Fig.~\ref{fig:Fermi_Surf_Distortion}. 

The charge longitudinal and SH conductivities are then obtained according
to $\sigma_{xx}=J_{x}/\mathcal{E}$ and $\sigma_{yz}^{z}=\mathcal{J}_{y}/\mathcal{E}$
with 
\begin{align}
J_{x} & =-g_{v}\int\frac{\Omega d^{2}\mathbf{k}}{(2\pi)^{2}}\left[\delta n_{\uparrow}(\mathbf{k})+\delta n_{\downarrow}(\mathbf{k})\right]v_{\mathbf{k},x}\,,\label{eq:j_x}\\
\mathcal{J}_{y} & =-g_{v}\int\frac{\Omega d^{2}\mathbf{k}}{(2\pi)^{2}}\left[\delta n_{\uparrow}(\mathbf{k})-\delta n_{\downarrow}(\mathbf{k})\right]v_{\mathbf{k},y}\,,\label{eq:j_y}
\end{align}
where $g_{v}=2$ is the valley degeneracy factor. As shown in Ref.~\cite{Ferreira14},
the zero-temperature conductivities admit an exact closed form 
\begin{equation}
\left.\begin{array}{cc}
\sigma_{xx}=2\,\epsilon\,\tau_{xx}\:,\quad\quad & \tau_{xx}\equiv\frac{\tau_{\parallel}^{\uparrow}}{1+\left(\frac{\tau_{\parallel}^{\uparrow}}{\tau_{\perp}^{\uparrow}}\right)^{2}}\:,\\
\sigma_{\textrm{SH}}=2\,\epsilon\,\tau_{yx}^{z}\:,\quad\quad & \tau_{yx}^{z}\equiv\frac{\tau_{\perp}^{\uparrow}}{1+\left(\frac{\tau_{\perp}^{\uparrow}}{\tau_{\parallel}^{\uparrow}}\right)^{2}}\:,
\end{array}\right.\label{eq:sigma_xx_xy}
\end{equation}
in terms of a parallel and perpendicular relaxation times, defined
by 
\begin{equation}
\frac{1}{\tau_{\parallel}^{\sigma}}=\frac{n\epsilon}{2\pi v^{2}}\int d\theta(1-\cos\theta)\,|T^{\sigma}(\theta)|^{2}\,,\label{eq:transport_rate}
\end{equation}
and 
\begin{equation}
\frac{1}{\tau_{\perp}^{\sigma}}=\frac{n\epsilon}{2\pi v^{2}}\int d\theta\sin\theta\,|T^{\sigma}(\theta)|^{2}\,,\label{eq:skew_rate}
\end{equation}
respectively \cite{Ferreira14}. Combining Eq.~(\ref{eq:T_matrix_eigen})
and Eqs.~\eqref{eq:transport_rate}-\eqref{eq:skew_rate} we find
after straightforward calculations 
\begin{align}
\frac{1}{\tau_{\parallel}^{\sigma}} & =\frac{n\epsilon}{4v^{2}}\left(\epsilon_{+}^{2}+\epsilon_{-}^{2}+\eta_{+}^{2}+\eta_{-}^{2}-\eta_{+}\eta_{-}-\epsilon_{+}\epsilon_{-}\right)\,,\label{eq:tau_p}\\
\frac{1}{\tau_{\perp}^{\sigma}} & =\frac{n\epsilon\sigma}{8v^{2}}\left(\epsilon_{-}\eta_{+}-\epsilon_{+}\eta_{-}\right)\,.\label{eq:tau_perp}
\end{align}
Inserting the above expressions in Eq.~(\ref{eq:sigma_xx_xy}) we
arrive at the desired result \begin{widetext} 
\begin{equation}
\sigma_{\textrm{SH}}=\frac{16e^{2}}{h}\left(\frac{v{}^{2}}{n}\right)\frac{\epsilon_{-}\eta_{+}-\epsilon_{+}\eta_{-}}{(\epsilon_{-}\eta_{+}-\epsilon_{+}\eta_{-})^{2}+4\left(\epsilon_{+}^{2}+\epsilon_{-}^{2}+\eta_{+}^{2}+\eta_{-}^{2}-\eta_{+}\eta_{-}-\epsilon_{+}\epsilon_{-}\right)^{2}}\,.\label{eq:SH_Boltzmann}
\end{equation}
\end{widetext} In Fig.~\eqref{fig:skew} we compare the skew scattering
contribution evaluated with Eq.~\eqref{eq:SH_Boltzmann} against
the self-consistent diagrammatic LRT expression {[}Eq.~\eqref{eq:FullSH}{]}.
We focus on the strong scattering regime, $|\Re\,g_{0}R^{2}\alpha_{3}|\gtrsim1$,
where the ``$Y$''-diagrams approximation breaks down (Sec.\,\ref{sub:VI-A-NonCroossing_Full_Diagramm}).
The two results are virtually indistinguishable. To better understand
this, we have performed an expansion in $|\Re\,g_{0}\,R^{2}(\alpha_{0}\pm\alpha_{3})|\ll1$
and found that the two expressions agree up to the third order. At
higher orders the expressions no longer coincide, but are numerically
very similar. The different mathematical structure of $\mathcal{S}(\epsilon)$
in the two approaches can be rationalized as follows. In the standard
Boltzmann description, electrons feel the scattering potential only
when they scatter off an impurity. Between two successive scatterings,
the electrons follow a straight trajectory determined by the classical
equation of motion~\cite{Rammer2}. Mathematically, this is expressed
by the fact that the matrix elements in Eq.~\eqref{eq:Fermi_Golden_Rule}
are evaluated over the eigenstates of the clean system. On the other
hand, in the Kubo formula disorder enters in two places, the self
energy and the vertex part, c.f. Sec.\,\ref{sec:II-METHODOLOGY}.
The latter gives the transport relaxation time, while the former corresponds
to dressing the bare eigenstates with disorder. We would like to mention
that the above picture is extremely clear when using the functional
approach~\cite{Atland}. There, the self energy is obtained from
the mean field solution (average disorder field seen by the electron)
and the vertex part is obtained by considering fluctuations around
this solution, \emph{i.e.}, fluctuations due to local scattering off
an impurity. Physically, this means that quantum mechanically the
electron's trajectory between successive scattering event is not a
straight line but it is affected by the background disorder field~\cite{Rammer2}.
As explained in detail in Sec.~\ref{sub:VI-A-NonCroossing_Full_Diagramm},
this is also the reason why the BTE does not capture per se the $\mathcal{Q}(\epsilon)$
term, as this is due to a cooperation of self energy and local scattering
effects.

\section{Beyond the noncrossing approximation\label{sec:VII-CROSSING}}

\begin{figure}[t!]
\centering{}\includegraphics[width=0.8\columnwidth]{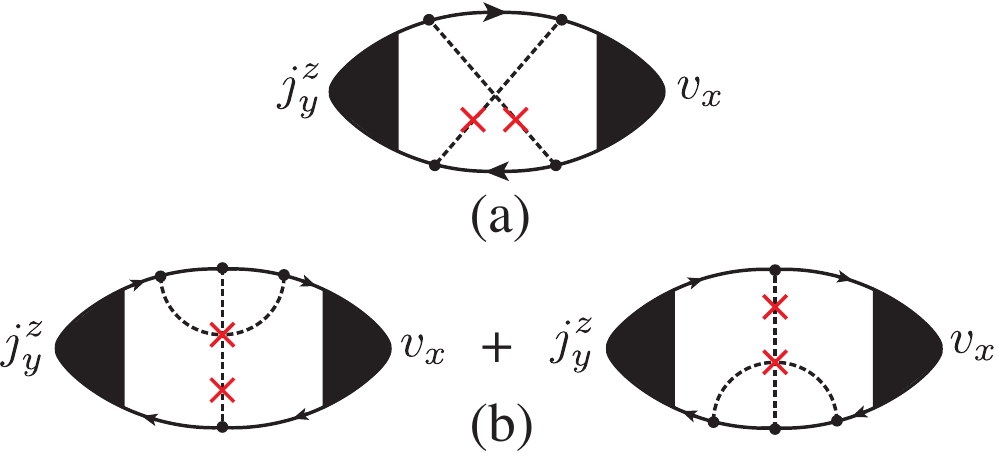} \caption{\label{fig:crossing}Diagrams with crossing impurities. (a) $X$ and
(b) $\Psi$ diagrams contributing to the SH conductivity. Here $\tilde{j}_{y}^{z}$
and $\tilde{v}_{x}$ are respectively the renormalized charge and
spin vertices. The dashed lines represent contractions of the impurity
potentials and the red $\times$ represent an $n$ insertion.}
 
\end{figure}

Crossing diagrams are usually associated with quantum interference
effects, and appear with an extra factor of smallness proportional
to $(k_{F}\,l)^{-1}$. Indeed, maximally crossed diagrams are responsible
for weak localization corrections~\cite{Rammer,Haug_Jauho_QuantumKinetics}.
Recently, Ado \emph{et al}.~\cite{Ado} showed in the context of
the AHE with massive Dirac fermions, that a specific subclass of crossing
diagrams also contribute to order order $(k_{F}\,l)^{0}$ to the transverse
conductivity. These diagrams (see Fig.~\ref{fig:crossing}), appearing
at fourth order in the impurity potential insertion, represent rare
events in which an electron skew scatters coherently off two impurities
located at a distance of the order of the Fermi wavelength $k_{F}\,|\mathbf{x}-\mathbf{x}'|\lesssim1$.
Being in the dilute regime, this is a rare event that nevertheless
gives rise to an anomalous contribution to the SH conductivity. As
it often happens in stochastic processes, rare events are associated
with big fluctuations above the average, and therefore can deeply
affect the value of observables. In this section, we evaluate the
contribution of the crossing diagrams to the SH conductivity shown
in the third line of Fig.~\ref{fig:general_exp}, first in the context
of the Gaussian approximation and then using the full $T$~matrix
formalism. 

\subsection{Gaussian model\label{sub:VI-A-Gaussian}}

\subsubsection{X-diagram\label{sub:VI-A1-Gauss_Xdiagram}}

The $X$ diagram is obtained by dressing the four point function with
four impurity potentials and two impurity density terms connecting
the retarded and advanced sectors, see Fig.~\ref{fig:crossing}.
Formally, one needs to take the disorder average of 
\begin{align}
\sigma_{X}^{\textrm{G}} & =2\int\prod_{n=0}^{4}\frac{d^{2}\mathbf{p}_{n}}{(2\pi)^{2}}\textrm{Tr}\Big\{\tilde{j}_{y}^{z}\,\mathcal{G}_{\mathbf{p}}^{R}\,\langle V(\mathbf{p-p_{1}})\,\mathcal{G}_{\mathbf{p}_{1}}^{R}\,V(\mathbf{p_{1}-p_{2}})\nonumber \\
 & \times\mathcal{G}_{\mathbf{p}_{2}}^{R}\,\tilde{v}_{x}\,\mathcal{G}_{\mathbf{p}_{2}}^{A}\,V(\mathbf{p_{2}-p_{3}})\,\mathcal{G}_{\mathbf{p}_{3}}^{A}\,V(\mathbf{p_{3}-p_{4}})\rangle_{\textrm{dis}}\,\mathcal{G}_{\mathbf{p}_{4}}^{A}\Big\},\label{eq:sigma_X}
\end{align}
where $\mathbf{p}_{0}\equiv\mathbf{p}$. The $X$ diagram is obtained
from the contraction 
\begin{align}
 & \langle V(\mathbf{p-p_{1}})V(\mathbf{p_{1}-p_{2}})V(\mathbf{p_{2}-p_{3}})V(\mathbf{p_{3}-p_{4}})\rangle_{\textrm{dis}}\nonumber \\
 & =\langle V(\mathbf{p-p_{1}})V(\mathbf{p_{2}-p_{3}})\rangle_{\textrm{dis}}\,\langle V(\mathbf{p_{1}-p_{2}})V(\mathbf{p_{3}-p_{4}})\rangle_{\textrm{dis}}.\label{eq:gausscorr}
\end{align}
In this way we obtain 
\begin{align}
\sigma_{X}^{\textrm{G}} & =2(n\,R^{4})^{2}\int\prod_{n=0}^{2}\frac{d^{2}\mathbf{p}_{n}}{(2\pi)^{2}}\,\textrm{Tr}\Big\{\left(M\mathcal{G}_{\mathbf{p}}^{A}\tilde{j}_{y}^{z}\,\mathcal{G}_{\mathbf{p}}^{R}M\right)\nonumber \\
 & \times\mathcal{G}_{\mathbf{p}_{1}}^{R}\,\left(M\mathcal{G}_{\mathbf{p}_{2}}^{R}\tilde{v}_{x}\,\mathcal{G}_{\mathbf{p}_{2}}^{A}M\right)\mathcal{G}_{\mathbf{Q}-\mathbf{p}_{1}}^{A}\Big\}\,,\label{eq:xbubble2}
\end{align}
where $\mathbf{Q}=\mathbf{p}+\mathbf{p}_{2}$ and the terms in round
brackets constitute the proper vertices. Due to the presence of $\mathbf{Q}$,
the momentum integrals do not factorize. Generally, the non factorization
of the integrals encodes correlation physics. In this case, it describes
an event in which the probability of an electron scattering off a
second, nearby impurity, depends on the scattering probability at
the first impurity. In order to evaluate this contribution, we follow
the approach of Ref.~\cite{Ado} and rewrite $\sigma_{X}^{\textrm{G}}$
in real space as

\begin{equation}
\sigma_{X}^{\textrm{G}}=2(nR^{4})^{2}\int_{\mathbf{r}}\textrm{Tr}\left\{ \Gamma_{y}^{z}(\mathbf{r})\mathcal{G}^{R}(-\mathbf{r})\Gamma_{x}(\mathbf{r})\mathcal{G}^{A}(-\mathbf{r})\right\} ,\label{eq:sigmaX_space}
\end{equation}
and 
\begin{align}
\Gamma_{y,\mathbf{r}}^{z} & =\int\frac{d^{2}\mathbf{p}}{(2\pi)^{2}}\left(M\,\mathcal{G}_{\mathbf{p}}^{A}\,\tilde{j}_{y}^{z}\,\mathcal{G}_{\mathbf{p}}^{R}\,M\right)e^{i\mathbf{p}\cdot\mathbf{r}}\,,\label{eq:Gamma_z_y}\\
\Gamma_{x,\mathbf{r}} & =\int\frac{d^{2}\mathbf{p}}{(2\pi)^{2}}\left(M\,\mathcal{G}_{\mathbf{p}}^{R}\,\tilde{v}_{x}\,\mathcal{G}_{\mathbf{p}}^{A}\,M\right)e^{i\mathbf{p}\cdot\mathbf{r}},\label{eq:Gamma_z_x}
\end{align}
where $\Gamma_{y,\mathbf{r}}^{z}$ and $\Gamma_{x,\mathbf{r}}$ are
respectively the Fourier transforms of the proper spin and charge
vertices. We look for contributions of the same order of the SH conductivity
evaluated in Eq.~\eqref{eq:SHdressed}. Since the $X$~conductivity
comes with a prefactor of $n^{2}$, this means that we need an additional
$1/n^{2}$ factor coming from the integrand. It is therefore enough
to keep the part of the renormalized vertex $\tilde{v}_{x}$ ($\tilde{j}_{y}^{z}$)
that is independent of $n$. According to Eq.~\eqref{eq:vertices}
we take: $\tilde{v}_{x}=F\,\gamma_{1}$ and $\tilde{j}_{y}^{z}=F/2\,\gamma_{13}$,
where $F=v+\delta v_{1}$. In order to evaluate the Fourier transform
of the proper vertices, we use the relation between the Dirac and
the Klein-Gordon propagator to write 
\begin{align}
\mathcal{G}_{\mathbf{p}}^{R/A} & =\left\{ (\epsilon\pm i\,n\,\eta)\gamma_{0}+n(m\mp i\,\bar{\eta})\gamma_{3}-iv\,\gamma^{j}\partial_{j}\right\} \label{eq:mixedG}\\
 & \times\frac{1}{(\epsilon\pm i\,n\,\eta)^{2}-n^{2}(m\mp i\,\bar{\eta})^{2}-v^{2}\,p^{2}},\nonumber 
\end{align}
to be understood in the operator sense. To lowest order in $n$, the
Fourier transform of the proper vertices reads 
\begin{align}
\Gamma_{x,\mathbf{r}} & =\frac{F}{n}\sum_{n=0}^{3}a_{n}(\mathbf{r})\gamma_{n}\,,\label{eq:xFourier}\\
\Gamma_{y,\mathbf{r}}^{z} & =\frac{F}{2n}\left\{ b_{0}(\mathbf{r})\gamma_{54}+b_{1}(\mathbf{r})\gamma_{23}+b_{2}(\mathbf{r})\gamma_{31}+b_{3}(\mathbf{r})\gamma_{12}\right\} ,\label{eq:yFourier}
\end{align}
where the space dependent coefficients $\{a_{n},b_{n}\}$ are defined
in Appendix~\ref{a:coeff}. Since the proper vertices already contain
a factor of $1/n$, the conductivity can be evaluated using the bare
Green's functions instead of the dressed ones~\cite{Ado}. The real
space form of the bare propagator reads 
\begin{equation}
\mathcal{G}_{0}^{R/A}(\mathbf{r})=\frac{\epsilon\,\gamma_{0}-i\,v\gamma^{j}\partial_{j}}{4v^{2}}\left[Y_{0}\left(\frac{\epsilon\,r}{v}\right)\mp i\,J_{0}\left(\frac{\epsilon\,r}{v}\right)\right],\label{eq:realBare}
\end{equation}
where $J_{0}$ and $Y_{0}$ are respectively Bessel functions of the
first and second kind, see Appendix~\ref{a:reals}. Using Eqs.~\eqref{eq:xFourier}-\eqref{eq:realBare}
in the expression for the conductivity, Eq.~\eqref{eq:sigmaX_space},
one finally obtains 
\begin{align}
\sigma_{X}^{\textrm{G}} & =-\frac{8e^{2}}{h}\,\frac{\alpha_{0}\,\alpha_{3}(\alpha_{0}^{2}-\alpha_{3}^{2})}{(\alpha_{0}^{2}+3\alpha_{3}^{2})^{2}}.\label{eq:sigmaX}
\end{align}
The $X$ diagram contribution to the SH conductivity has the basic
symmetry of the semiclassical skew scattering, namely it vanishes
when $|\alpha_{0}|=|\alpha_{3}|$, see Eq.~\eqref{eq:yweak}.

\begin{figure}[t!]
\centering{}\includegraphics[width=1\columnwidth]{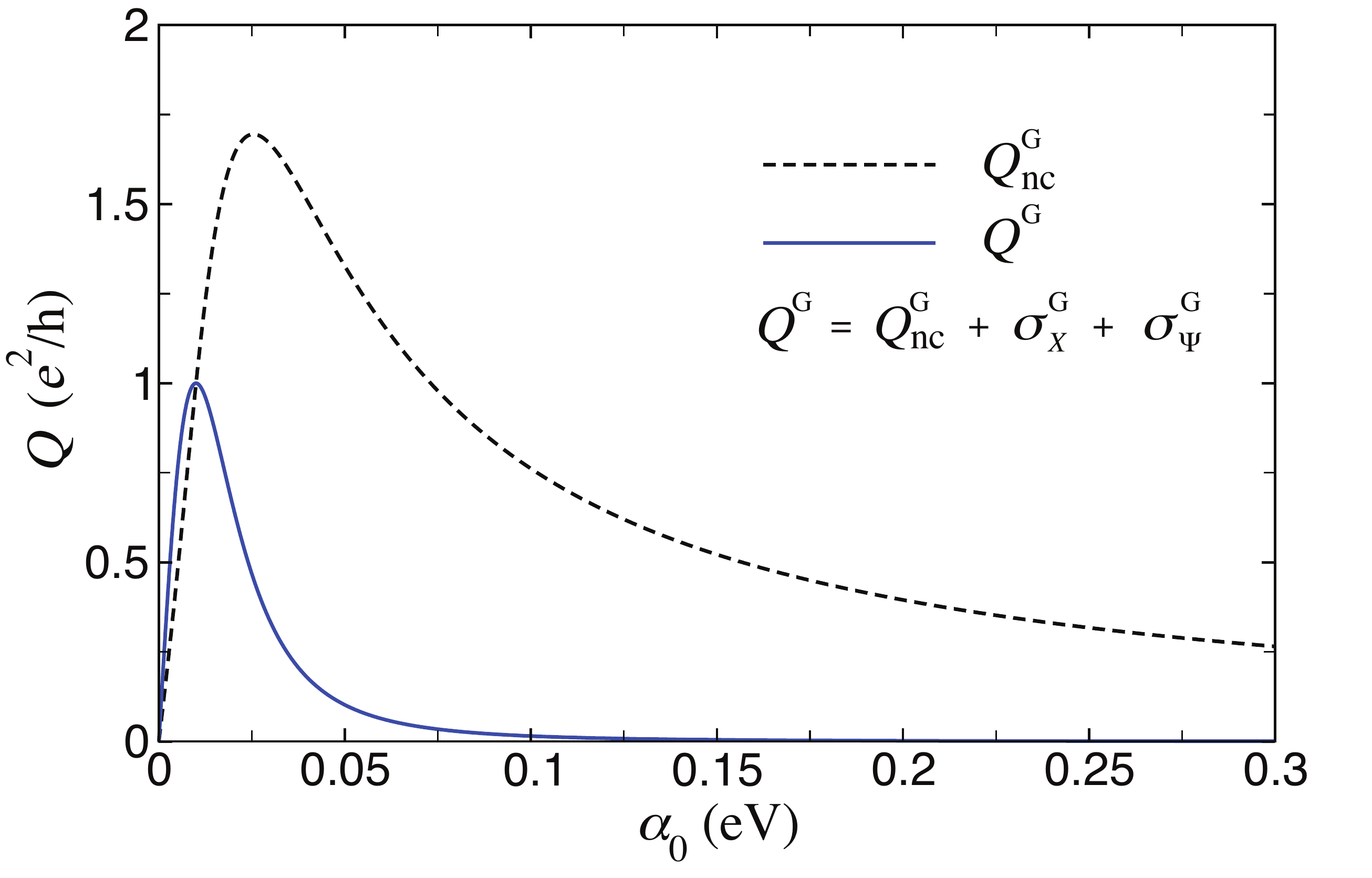} \caption{\label{fig:SHtotal}Total anomalous contribution to the SH conductivity
in the Gaussian approximation.\textbf{ }The noncrossing result is
shown for comparison. The calculation has $\alpha_{3}=10$~meV. }
\end{figure}

\subsubsection{$\Psi$-diagrams\label{sub:VI-A2-Gauss_Psidiagram}}

We now evaluate the two $\Psi$ diagrams of Fig.~\ref{fig:crossing}(b).
The evaluation closely follow that of the $X$ diagram, therefore
we only highlights the main steps. For the first diagram, we find
in momentum space 
\begin{align}
\sigma_{\Psi,\textrm{I}}^{\textrm{G }} & =2\,(nR^{4})^{2}\,\int\prod_{n=0}^{2}\frac{d^{2}\mathbf{p}_{n}}{(2\pi)^{2}}\textrm{Tr}\Big\{\left(M\,\mathcal{G}_{\mathbf{p}}^{A}\,\tilde{j}_{y}^{z}\,\mathcal{G}_{\mathbf{p}}^{R}\,M\right)\,\nonumber \\
 & \times\mathcal{G}_{\mathbf{k}+\mathbf{p}_{1}}^{R}\,M\,\mathcal{G}_{\mathbf{p}_{1}}^{R}\,M\,\left(\mathcal{G}_{\mathbf{p}_{2}}^{R}\,\tilde{v}_{x}\,\mathcal{G}_{\mathbf{p}_{2}}^{A}\right)\Big\}\,,\label{eq:psi1}
\end{align}
where $\mathbf{k}=\mathbf{p}-\mathbf{p}_{2}$ and as before $\mathbf{p}_{0}\equiv\mathbf{p}$.
The second diagram can be obtained from the first by taking the conjugate
of $\sigma_{\Psi,\textrm{I}}^{\textrm{G }}$. Also in this case, it
is convenient to move to real space. The $\Psi$ diagrams contribution
to the conductivity then reads 
\begin{align}
\sigma_{\Psi}^{\textrm{G}} & =2\,(nR^{4})^{2}\int_{\mathbf{r}}\,\textrm{Tr}\left\{ \Gamma_{y}^{z}(\mathbf{r})\,\mathcal{G}^{R}(-\mathbf{r})\,M\,\mathcal{G}^{R}(\mathbf{r})\,M\,\chi_{x}(-\mathbf{r})\right.\nonumber \\
 & \left.+\:\Gamma_{y}^{z}(\mathbf{r})\,\chi_{x}(-\mathbf{r})\,M\,\mathcal{G}^{A}(\mathbf{r})\,M\,\mathcal{G}^{A}(-\mathbf{r})\right\} \,,\label{eq:sigma_Psi_space}
\end{align}
where 
\begin{align}
\chi_{x}(\mathbf{r}) & =\int\frac{d^{2}\mathbf{p}}{(2\pi)^{2}}\left(\mathcal{G}_{\mathbf{p}_{2}}^{R}\,\tilde{v}_{x}\,\mathcal{G}_{\mathbf{p}_{2}}^{A}\right)e^{i\mathbf{p}\cdot\mathbf{r}},\label{eq:propvert}
\end{align}
and $\Gamma_{y,\mathbf{r}}^{z}$ is the same as in Eq.~\eqref{eq:yFourier}.
To lowest order in $n$, the evaluation of the Fourier transform of
the proper charge vertex yields 
\begin{equation}
\chi_{x}(\mathbf{r})=\frac{F}{n}\sum_{n=0}^{2}c_{n}(\mathbf{r})\gamma_{n}\,,\label{eq:propvert1}
\end{equation}
where the explicit expression for the coefficients are given in Appendix~\ref{a:coeff}.
Performing the real space integral we find $\sigma_{\Psi}=0$. A similar
result was found by Ado \emph{et al}.~\cite{Ado} in the context
of the AHE, where the authors also stressed that the vanishing of
$\sigma_{\Psi}$ is due to the minimal form of the model. One could
argue that this is the case also in our model. However, and as we
show below, we find that the vanishing of $\sigma_{\Psi}$ is again
an artifact of the Gaussian approximation.

Summing the non-crossing, $\Psi$ and $X$ contributions we finally
arrive at 
\begin{equation}
\sigma_{\textrm{SH}}^{\textrm{G}}=Q_{\textrm{nc}}^{\textrm{G}}+\sigma_{X}^{\textrm{G}}+\sigma_{\Psi}^{\textrm{G}}=\frac{16e^{2}}{h}\,\frac{\alpha_{0}\,\alpha_{3}^{3}}{\left(\alpha_{0}^{2}+3\,\alpha_{3}^{2}\right)^{2}}.\label{eq:totSH}
\end{equation}
The total SH conductivity in the Gaussian approximation is plotted
in Fig.~\ref{fig:SHtotal} as a function of $\alpha_{0}$. We see
that the crossing diagrams drastically reduce the value of the SH
conductivity with respect to the noncrossing approximation.

\subsection{$T$~Matrix evaluation\label{sub:VI-B-Tmatrix}}

As we have explained before, the lack of an energy dependence in the
expression for the SH conductivity is an artifact of the Gaussian
approximation. In Ref.~\cite{Milletari_1} we have showed that using
the full $T$~matrix, one obtains indeed a Fermi energy dependence
of the quantum anomalous SH conductivity, $\mathcal{Q}$. Obtaining
the correct expression for the energy dependence is crucial for assessing
the crossover between semiclassical to quantum anomalous spin transport.

\begin{figure}[t!]
\centering{}\includegraphics[width=1\columnwidth]{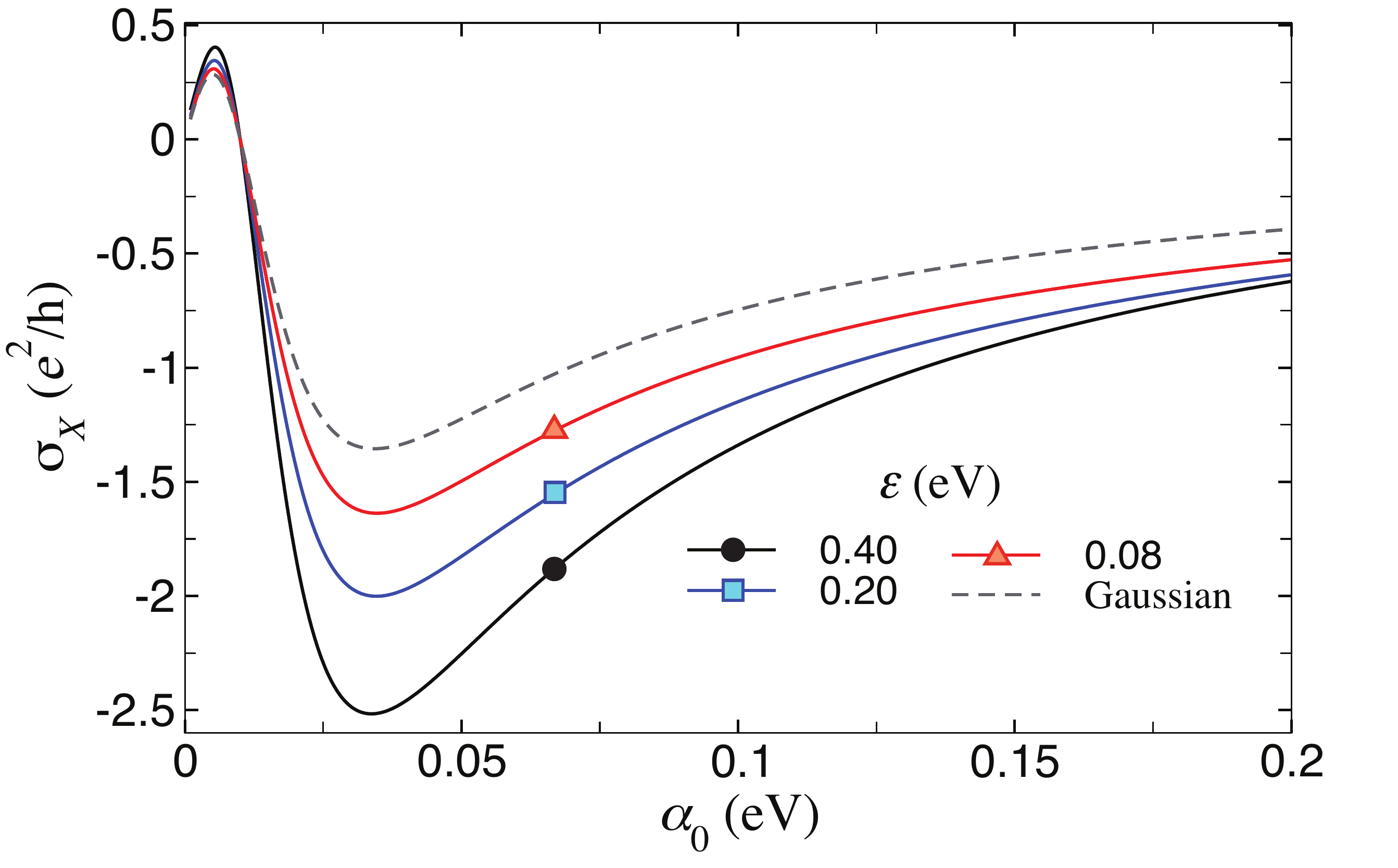} \caption{$X$~conductivity in the full $T$~matrix approach. The $X$~conductivity
is plotted as a function of the scalar disorder potential $\alpha_{0}$
for different values of the Fermi energy. As a comparison, we also
plot the Gaussian value, formally recovered in the $\epsilon\to0$
limit. Note that $\sigma_{X}$ goes to zero at higher values of $\alpha_{0}$,
in agreement with the Unitary limit result. The other parameters used
in the plot are: $\alpha_{3}=10$~meV, and $R=4\,{\rm nm}$. }
\label{fig:SX} 
\end{figure}

Two main differences arise when performing the calculation of crossing
diagrams within the $T$~matrix approach\textit{\emph{. First, }}the
$b$ coefficient of the renormalized vertex in Eq.~\eqref{eq:ladder}
now contains an extra term that is independent of $n$, see Eq.~\eqref{eq:Tladder}.
As we showed in Sec.\,\ref{sub:VI-A-NonCroossing_Full_Diagramm},
in the noncrossing approximation this additional vertex part is responsible
for the semiclassical skew scattering contribution, introducing an
effective spin-spin interaction~\cite{Milletari_1}. Mathematically,
the additional vertex part modifies the structure of the proper vertices
in Eqs.~\eqref{eq:xFourier}-\eqref{eq:yFourier}. For each vertex,
we now have two contributions, namely, $\Gamma_{x,\mathbf{r}}=\Gamma_{x,\mathbf{r}}^{1}+\Gamma_{x,\mathbf{r}}^{2}$,
where 
\begin{align}
\Gamma_{x,\mathbf{r}}^{1} & =\frac{1}{n}\sum_{n=0}^{3}\bar{a}_{n}(\mathbf{r})\gamma_{n}\,,\label{eq:xFourier_TM1}\\
\Gamma_{x,\mathbf{r}}^{2} & =\frac{1}{n}\left\{ \bar{b}_{0}(\mathbf{r})\gamma_{54}+\bar{b}_{1}(\mathbf{r})\gamma_{23}+\bar{b}_{2}(\mathbf{r})\gamma_{31}+\bar{b}_{3}(\mathbf{r})\gamma_{12}\right\} ,\label{eq:xFourier_TM2}
\end{align}
and $\Gamma_{y,\mathbf{r}}^{z}=\Gamma_{y,\mathbf{r}}^{z,1}+\Gamma_{y,\mathbf{r}}^{z,2}$,
where now the coefficients are interchanged with respect to the charge
vertex, \emph{i.e.}, 
\begin{align}
\Gamma_{y,\mathbf{r}}^{z,1} & =\frac{1}{n}\sum_{n=0}^{3}\bar{b}_{n}(\mathbf{r})\gamma_{n}\,,\label{eq:yFourier_TM1}\\
\Gamma_{y,\mathbf{r}}^{z,2} & =\frac{1}{n}\left\{ \bar{a}_{0}(\mathbf{r})\gamma_{54}+\bar{a}_{1}(\mathbf{r})\gamma_{23}+\bar{a}_{2}(\mathbf{r})\gamma_{31}+\bar{a}_{3}(\mathbf{r})\gamma_{12}\right\} \,.
\end{align}

Secondly, the single potential insertions in the crossing bubbles
of Fig.~\ref{fig:crossing} need to be replaced with the $T$~matrix,
just like we did in Ref.~\cite{Milletari_1} for the ladder diagram.
This last step ensures the resummation of all topologically equivalent
diagrams. Apart from these differences, the evaluation of the crossing
contribution follows the steps of Sec.\,\ref{sub:VI-A-NonCroossing_Full_Diagramm}.
After a lengthy calculation we obtain the Fermi energy dependence
of the $X$ diagram. This is shown in Fig.~\ref{fig:SX}, where we
also plot the Gaussian result, Eq.~\eqref{eq:sigmaX}, as a guide.
The magnitude of the $X$ diagram contribution increases with Fermi
energy, as expected for a skew scattering contribution. We note that
the Gaussian result is recovered from the self-consistent one in the
formal $\epsilon\to0$ limit.

\begin{figure}[t!]
\centering{}\includegraphics[width=1\columnwidth]{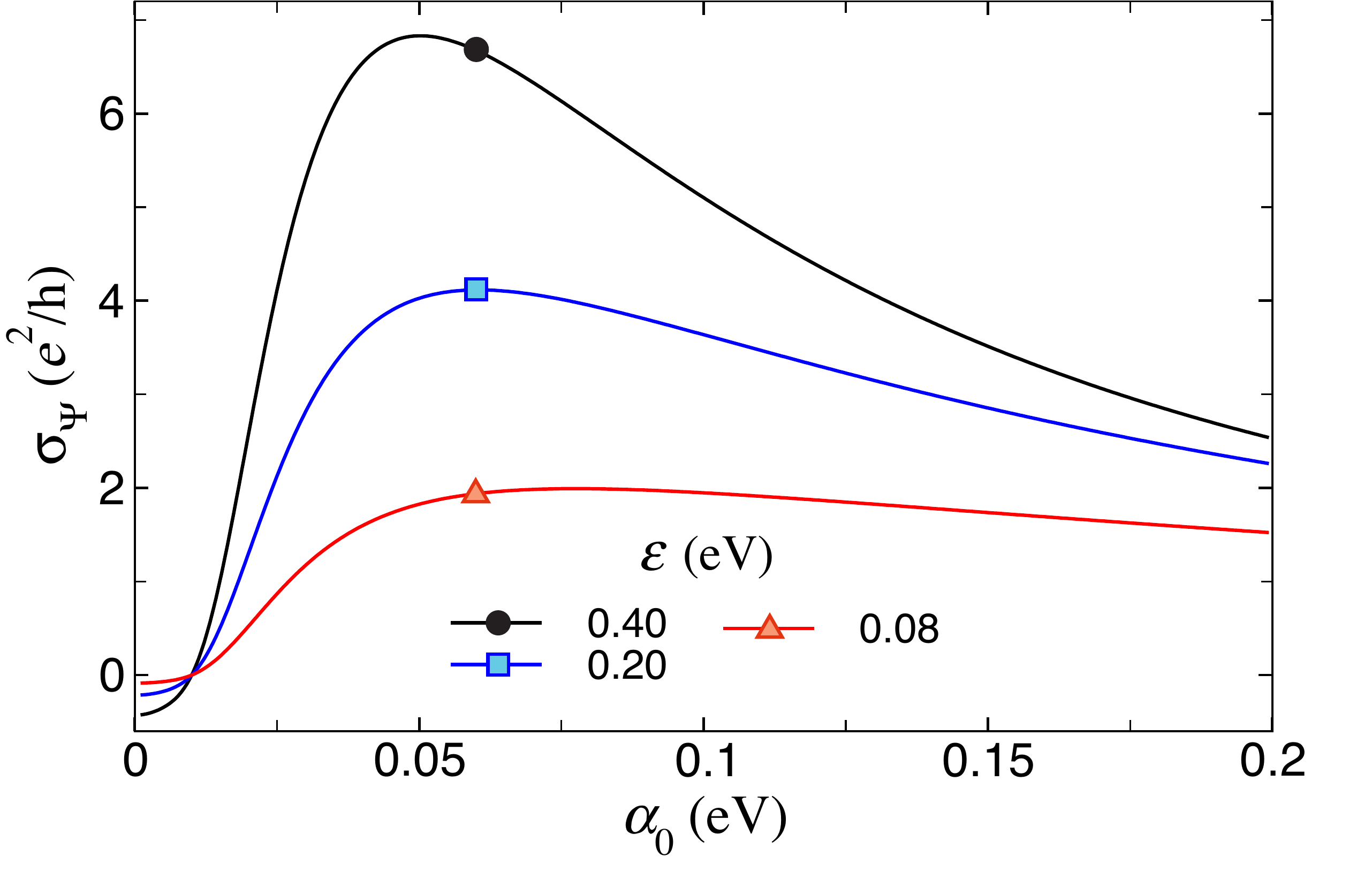} \caption{\label{fig:SP}$\Psi$~conductivity in the full $T$~matrix approach.
The $\Psi$~conductivity is plotted as a function of the scalar potential
strength $\alpha_{0}$ for different values of the Fermi energy. Note
that $\sigma_{\Psi}\to0$ as $\epsilon\to0$. At higher values of
$\alpha_{0}$, $\sigma_{\Psi}$ also goes to zero in agreement with
the unitary limit result. The other values used in the plot are: $\alpha_{3}=0.01$~eV,
and $R=4$~nm.}
\end{figure}

Most importantly, we find that the $\Psi$ diagrams give a \emph{finite}
contribution away from the Dirac point. This is shown in Fig.~\ref{fig:SP},
where the $\Psi$~contribution to the SH conductivity is plotted
for different values of the Fermi energy. Strikingly, the $\Psi$
diagram is found to result in the dominant contribution to $\mathcal{Q}$
over a wide range of parameters, whereas in the Gaussian approximation
this term plays no role. The enhanced quantum coherent contribution
away from the Dirac point is one of the central findings of the current
work. Comparing the $\Psi$ contribution with the semiclassical skew
scattering, it is easy to recognize several similarities: $\sigma_{\Psi}$
increases as the Fermi energy is increased. Independently of the value
of $\epsilon$, $\sigma_{\Psi}$ is zero if $|\alpha_{0}|=|\alpha_{3}|$
and finally, there is a negative contribution for $\alpha_{0}<\alpha_{3}$.
This term is dominant at high Fermi energy, and therefore it can compete
with the semiclassical skew scattering term, favoring in this way
the crossover to the quantum anomalous dominated SH regime recently
discovered in Ref.~\cite{Milletari_1}.

\section{Conclusions\label{sec:VII-Conclusions}}

In this work we have presented a detailed analysis of the spin Hall
effect in systems of 2D massless Dirac fermions subject to extrinsic
spin--orbit interactions with origin in short-range impurities. We
have shown that for meaningful scattering potential models---characterized
by multiple terms in the low-energy theory, or that are generally
in the strong scattering regime---the often used Gaussian-type approximations
fail to give the correct answers, especially by predicting a spin
Hall response function independent of the Fermi energy. In order to
overcome these limitations, we introduced \textcolor{black}{a self-consistent
quantum diagrammatic approach, in which self energy operators and
vertex corrections in the Kubo--Streda formula are evaluated at the
$T$~matrix level. The extended diagrammatic technique allows us
to treat skew scattering and anomalous contributions on equal footing,
and at all orders in the potential strength, by resumming all moments
of the disorder distribution. In the noncrossing approximation, we
find that the spin Hall conductivity $\sigma_{\textrm{SH}}$ depends
on the Fermi energy in a sensible way. The (semiclassical) skew scattering
contribution $\sigma_{\textrm{SH}}\propto n^{-1}$ is shown to quantitatively
agree with Boltzmann transport theory over a wide range of parameters.
To our knowledge, this is the first work showing the equivalence between
semiclassical and quantum linear response approaches in the strong
scattering regime in the presence of spin--orbit coupling. Furthermore,
by treating skew scattering and quantum side jump processes on equal
footing, we show that these mechanisms are in general correlated even
at the noncrossing level. This suggests that previous treatments,
assuming independent contributions from skew scattering and quantum
side-jump events, are not generally valid. Finally, we have evaluated
the contribution of an important subclass of crossing diagrams encoding
quantum interference corrections, and showed they contribute to the
anomalous component of the spin Hall conductivity. We have shown that
also in this case the Gaussian approximation fails to give the correct
picture, especially by predicting the vanishing of a subclass of crossing
diagrams. Within the extended $T$~matrix formalism we not only show
that this class of diagrams is non vanishing, but it dominates the
spin Hall conductivity in experimentally relevant parameter regions.
Most importantly, these quantum coherent corrections seem to favor
the crossover to the anomalous regime recently discussed by the authors
in Ref.~\cite{Milletari_1}, opening the exciting prospect of detecting
signatures of quantum coherent spin Hall phenomena in nonlocal transport
experiments.}

\section{Acknowledgements\label{sec:VIII-Acknowledgements}}

M.M. thanks R. Raimondi and G. Vignale for stimulating discussions.
M.M. acknowledges support from the Singapore National Research Foundation
under its fellowship program (NRF Award No. NRF-NRFF2012-01). A.F.
gratefully acknowledges the financial support of the Royal Society
(U.K.) through a Royal Society University Research Fellowship.

\appendix
\begin{widetext}

\section{Functions appearing in the self consistent approach\label{a:SELFCONS}}

\begin{equation}
f_{a}(\eta_{+},\eta_{-},\epsilon_{+},\epsilon_{-})=\frac{(\eta_{+}+\eta_{-})(\epsilon_{+}\epsilon_{-}+\eta_{+}\eta_{-})-\pi(\eta_{+}-\eta_{-})(\epsilon_{+}\eta_{-}-\eta_{+}\epsilon_{-})}{4\pi v^{2}(\eta_{+}+\eta_{-})}\,.\label{s:fa}
\end{equation}
\begin{equation}
f_{b}(\eta_{+},\eta_{-},\epsilon_{+},\epsilon_{-})=\frac{(\eta_{+}+\eta_{-})(\epsilon_{+}\eta_{-}-\eta_{+}\epsilon_{-})+\pi(\eta_{+}-\eta_{-})(\epsilon_{+}\epsilon_{-}+\eta_{+}\eta_{-})}{4\pi v^{2}(\eta_{+}+\eta_{-})}\,.
\end{equation}

\begin{align}
\delta v_{10} & =v\frac{4\,v^{2}\,\epsilon\,(\eta_{+}+\eta_{-})(\epsilon_{+}\epsilon_{-}+\eta_{+}\eta_{-})-\epsilon^{2}\left(\eta_{+}^{2}+\epsilon_{+}^{2}\right)\left(\eta_{-}^{2}+\epsilon_{-}^{2}\right)}{\epsilon^{2}\left(\eta_{+}^{2}+\epsilon_{+}^{2}\right)\left(\eta_{-}^{2}+\epsilon_{-}^{2}\right)-8\,\epsilon\,v^{2}\,(\eta_{+}+\eta_{-})(\epsilon_{+}\epsilon_{-}+\eta_{+}\eta_{-})+16\,v^{4}\,(\eta_{+}+\eta_{-})^{2}}\,.\label{eq:Tvert}
\end{align}
\[
\delta v_{20}=v\,\frac{4\,\epsilon\,v^{2}\,(\eta_{+}+\eta_{-})(\epsilon_{-}\eta_{+}-\epsilon_{+}\eta_{-})}{\epsilon^{2}\left(\eta_{+}^{2}+\epsilon_{+}^{2}\right)\left(\eta_{-}^{2}+\epsilon_{-}^{2}\right)-8\,\epsilon\,v^{2}\,(\eta_{+}+\eta_{-})(\epsilon_{+}\epsilon_{-}+\eta_{+}\eta_{-})+16\,v^{4}\,(\eta_{+}+\eta_{-})^{2}}\,.
\]

\begin{align}
\delta v_{11} & =\frac{v}{\pi}4\,v^{2}(\eta_{+}+\eta_{-})\Big\{16v^{4}(\eta_{+}-\eta_{-})^{2}[\pi(\eta_{+}+\eta_{-})(\eta_{+}\epsilon_{-}-\eta_{-}\epsilon_{+})+(\eta_{+}+\eta_{-})(\eta_{+}\eta_{-}+\epsilon_{+}\epsilon_{-})]-8\epsilon\pi v^{2}(\eta_{+}+\eta_{-})^{2}\nonumber \\
 & \times\left(\eta_{+}^{2}+\epsilon_{+}^{2}\right)\left(\eta_{-}^{2}+\epsilon_{-}^{2}\right)+\epsilon^{2}\left(\eta_{+}^{2}+\epsilon_{+}^{2}\right)\left(\eta_{-}^{2}+\epsilon_{-}^{2}\right)[(\eta_{+}+\eta_{-})(\eta_{+}\eta_{-}+\epsilon_{+}\epsilon_{-})-\pi(\eta_{+}+\eta_{-})(\eta_{+}\epsilon_{-}-\eta_{-}\epsilon_{+})]\Big\}\nonumber \\
 & /\left[\epsilon^{2}\left(\eta_{+}^{2}+\epsilon_{+}^{2}\right)\left(\eta_{-}^{2}+\epsilon_{-}^{2}\right)-8\epsilon\,v^{2}(\eta_{+}+\eta_{-})(\epsilon_{+}\epsilon_{-}+\eta_{+}\eta_{-})+16v^{4}(\eta_{+}+\eta_{-})^{2}\right]^{2}\,.
\end{align}

\begin{align}
\delta v_{22} & =\frac{v}{\pi}4\,v^{2}(\eta_{+}+\eta_{-})\Big\{16v^{4}(\eta_{+}-\eta_{-})^{2}[\pi(\eta_{+}-\eta_{-})(\eta_{+}\eta_{-}+\epsilon_{+}\epsilon_{-})+(\eta_{+}^{2}+\eta_{-}^{2})(\eta_{+}\epsilon_{-}-\eta_{-}\epsilon_{+})]-8\epsilon\pi v^{2}(\eta_{+}-\eta_{-})\nonumber \\
 & \times\left(\eta_{+}^{2}+\epsilon_{+}^{2}\right)\left(\eta_{-}^{2}+\epsilon_{-}^{2}\right)+\epsilon^{2}\left(\eta_{+}^{2}+\epsilon_{+}^{2}\right)\left(\eta_{-}^{2}+\epsilon_{-}^{2}\right)[(\eta_{+}+\eta_{-})(\eta_{+}\epsilon_{-}-\eta_{-}\epsilon_{+})+\pi(\eta_{+}-\eta_{-})(\eta_{+}\eta_{-}+\epsilon_{+}\epsilon_{-})]\Big\}\nonumber \\
 & /\left[\epsilon^{2}\left(\eta_{+}^{2}+\epsilon_{+}^{2}\right)\left(\eta_{-}^{2}+\epsilon_{-}^{2}\right)-8\epsilon\,v^{2}(\eta_{+}+\eta_{-})(\epsilon_{+}\epsilon_{-}+\eta_{+}\eta_{-})+16v^{4}(\eta_{+}+\eta_{-})^{2}\right]^{2}\,.
\end{align}

\section{Real Space form of the propagator\label{a:reals}}

In order to obtain the real space form of the bare propagator, one
starts from the operator relation 
\begin{equation}
\mathcal{G}_{0}^{R/A}(\mathbf{r})=\left(\epsilon\,\gamma_{0}-i\,v\,\gamma^{i}\partial_{i}\right)\,\mathcal{F}\left(\frac{1}{\epsilon^{2}-v^{2}\,k^{2}\pm i\,0^{+}}\right),\label{eq:DiracKG}
\end{equation}
where $\mathcal{F}[.]$ stands for Fourier transform. Explicitly,
\begin{equation}
\mathcal{F}[\left(\frac{1}{\epsilon^{2}-v^{2}\,k^{2}\pm i\,0^{+}}\right)=-\frac{1}{2\pi\,v^{2}}\,K_{0}\left(\mp\,i\frac{\epsilon\,r}{v}\right)=\frac{1}{4\,v^{2}}\left[Y_{0}\left(\frac{\epsilon\,r}{v}\right)\mp\,i\,J_{0}\left(\frac{\epsilon\,r}{v}\right)\right],\label{eq:FT}
\end{equation}
where $r\equiv|\mathbf{r}|$. In the last step we have separated real
and imaginary part of the modified Bessel function $K_{0}(z)$ using
the identity 
\begin{equation}
-\frac{2}{\pi}\,K_{\nu}(z)\,e^{-i\,\pi\nu/2}=Y_{\nu}(\imath\,z)-i\,J_{\nu}(i\,z),\label{eq:iden}
\end{equation}
where $J_{\nu}$ and $Y_{\nu}$ are Bessel functions of the first
and second kind, respectively (see e.g., \url{http://dlmf.nist.gov/10.27}).
The real space propagator then reads 
\begin{equation}
\mathcal{G}_{0}^{R/A}(\mathbf{r})=\frac{\epsilon}{4\,v^{2}}\left\{ \gamma_{0}\,\left[Y_{0}\left(\frac{\epsilon\,r}{v}\right)\mp\,i\,J_{0}\left(\frac{\epsilon\,r}{v}\right)\right]+i\left(\gamma_{1}\,\hat{x}_{1}+\gamma_{2}\,\hat{x}_{2}\right)\left[Y_{1}\left(\frac{\epsilon\,r}{v}\right)\mp\,i\,J_{1}\left(\frac{\epsilon\,r}{v}\right)\right]\right\} ,\label{eq:G_exp}
\end{equation}
where $\hat{x}_{i}$ is the unit vector in two dimensions.

\section{Coefficients appearing in the proper vertices I: Gaussian\label{a:coeff}}

Here we list the coefficients appearing in the definition of the proper
vertices. The $J_{\nu}$ are Bessel functions of the first kind. 
\begin{align}
a_{0}(\mathbf{r}) & =\frac{i\,\cos(\theta)J_{1}\left(\frac{\epsilon\,r}{v}\right)}{R^{4}}\,.\label{eq:acoeff}\\
a_{1}(\mathbf{r}) & =\frac{\left(\alpha_{0}^{2}-\alpha_{3}^{2}\right)}{2\,R^{4}\left(\alpha_{0}^{2}+\alpha_{3}^{2}\right)}\left\{ J_{0}\left(\frac{\epsilon\,r}{v}\right)+\cos(2\theta)\frac{\left[\epsilon\,r\,J_{0}\left(\frac{\epsilon\,r}{v}\right)-2\,v\,J_{1}\left(\frac{\epsilon\,r}{v}\right)-\epsilon\,r\,J_{2}\left(\frac{\epsilon\,r}{v}\right)\right]}{2\,\epsilon\,r}\right\} \,.\\
a_{2}(\mathbf{r}) & =\frac{-\left(\alpha_{0}^{2}-\alpha_{3}^{2}\right)\,\sin(2\theta)}{R^{4}\left(\alpha_{0}^{2}+\alpha_{3}^{2}\right)\,\epsilon}\left\{ \frac{\epsilon\,r\,J_{0}\left(\frac{\epsilon\,r}{v}\right)-2\,v\,J_{1}\left(\frac{\epsilon\,r}{v}\right)-\epsilon\,r\,J_{2}\left(\frac{\epsilon\,r}{v}\right)}{4\,r}\right\} \,.\\
a_{3}(\mathbf{r}) & =\frac{2\,i\,\alpha_{0}\,\alpha_{3}\,\cos(\theta)\,J_{1}\left(\frac{\epsilon\,r}{v}\right)}{R^{4}\,\left(\alpha_{0}^{2}+\alpha_{3}^{2}\right)}.
\end{align}
\\

\begin{align}
b_{0}(\mathbf{r}) & =\frac{i\,\sin(\theta)\,J_{1}\left(\frac{\epsilon\,r}{v}\right)}{R^{4}}\,.\label{eq:bcoeff}\\
b_{1}(\mathbf{r}) & =\frac{-(\alpha_{0}^{2}-\alpha_{3}^{2})\sin(2\theta)}{R^{4}\left(\alpha_{0}^{2}+\alpha_{3}^{2}\right)\,\epsilon}\left\{ \frac{\epsilon\,rJ_{0}\left(\frac{\epsilon\,r}{v}\right)-2\,v\,J_{1}\left(\frac{\epsilon\,r}{v}\right)-\epsilon\,r\,J_{2}\left(\frac{\epsilon\,r}{v}\right)}{4r}\right\} \,.\\
b_{2}(\mathbf{r}) & =\frac{\alpha_{0}^{2}-\alpha_{3}^{2}}{2\,R^{4}\left(\alpha_{0}^{2}+\alpha_{3}^{2}\right)}\left\{ J_{0}\left(\frac{\epsilon\,r}{v}\right)-\cos(2\theta)\frac{\left[\epsilon\,rJ_{0}\left(\frac{\epsilon\,r}{v}\right)-2\,v\,J_{1}\left(\frac{\epsilon\,r}{v}\right)-\epsilon\,rJ_{2}\left(\frac{\epsilon\,r}{v}\right)\right]}{2\,\epsilon\,r}\right\} \,.\\
b_{3}(\mathbf{r}) & =\frac{2\,i\,\alpha_{0}\,\alpha_{3}\,\sin(\theta)\,J_{1}\left(\frac{\epsilon\,r}{v}\right)}{R^{4}\left(\alpha_{0}^{2}+\alpha_{3}^{2}\right)}\,.
\end{align}
\\

\begin{align}
c_{0}(\mathbf{r}) & =\frac{i\,\cos(\theta)\,J_{1}\left(\frac{\epsilon\,r}{v}\right)}{R^{4}\left(\alpha_{0}^{2}+\alpha_{3}^{2}\right)}\,.\label{eq:ccoeff}\\
c_{1}(\mathbf{r}) & =\frac{1}{2\,R^{4}\left(\alpha_{0}^{2}+\alpha_{3}^{2}\right)}\left\{ J_{0}\left(\frac{\epsilon\,r}{v}\right)+\cos(2\theta)\frac{\left[\epsilon\,rJ_{0}\left(\frac{\epsilon\,r}{v}\right)-2\,v\,J_{1}\left(\frac{\epsilon\,r}{v}\right)-\epsilon\,r\,J_{2}\left(\frac{\epsilon\,r}{v}\right)\right]}{2\,\epsilon\,r}\right\} \,.\\
c_{2}(\mathbf{r}) & =\frac{1}{R^{4}\left(\alpha_{0}^{2}+\alpha_{3}^{2}\right)\,\epsilon}\left\{ \sin(2\theta)\frac{\left[\epsilon\,rJ_{0}\left(\frac{\epsilon\,r}{v}\right)-2\,v\,J_{1}\left(\frac{\epsilon\,r}{v}\right)-\epsilon\,r\,J_{2}\left(\frac{\epsilon\,r}{v}\right)\right]}{4\,r}\right\} \,.
\end{align}

\section{Coefficients appearing in the proper vertices II: T-Matrix}

\label{a:coeff1}

\begin{align}
\bar{a}_{0}(\mathbf{r}) & =\frac{i\,(v+\delta v_{10})\left(\eta_{+}^{2}+\eta_{-}^{2}+\epsilon_{+}^{2}+\epsilon_{-}^{2}\right)\,\epsilon\cos(\theta)J_{1}\left(\frac{\epsilon\,r}{v}\right)}{4\,v^{2}(\eta_{+}+\eta_{-})}\,.\label{eq:abcoeff}\\
\bar{a}_{1}(\mathbf{r}) & =\frac{(v+\delta v_{10})(\eta_{+}\,\eta_{-}+\epsilon_{+}\,\epsilon_{-})}{4\,\epsilon\,v^{2}(\eta_{+}+\eta_{-})}\left\{ \epsilon^{2}J_{0}\left(\frac{\epsilon\,r}{v}\right)+\frac{\epsilon\,\cos(2\theta)\left[\epsilon\,r\,J_{0}\left(\frac{\epsilon\,r}{v}\right)-2\,v\,J_{1}\left(\frac{\epsilon\,r}{v}\right)-\epsilon\,r\,J_{2}\left(\frac{\epsilon\,r}{v}\right)\right]}{2r}\right\} \\
 & +\frac{\delta v_{20}(\eta_{+}\,\epsilon_{-}-\eta_{-}\,\epsilon_{+})}{4\,\epsilon\,v^{2}\,(\eta_{+}+\eta_{-})}\left\{ \frac{\epsilon\,\cos(2\theta)\left[\epsilon\,rJ_{0}\left(\frac{\epsilon\,r}{v}\right)-2\,v\,J_{1}\left(\frac{\epsilon\,r}{v}\right)-\epsilon\,r\,J_{2}\left(\frac{\epsilon\,r}{v}\right)\right]}{2\,r}-\epsilon^{2}J_{0}\left(\frac{\epsilon\,r}{v}\right)\right\} \,.\nonumber \\
\bar{a}_{2}(\mathbf{r}) & =\frac{\delta v_{20}(\eta_{+}\,\epsilon_{-}-\eta_{-}\,\epsilon_{+})+(v+\delta v_{10})(\eta_{+}\,\eta_{-}+\epsilon_{+}\epsilon_{-})}{2\,v^{2}\,(\eta_{+}+\eta_{-})}\left\{ \frac{\sin(2\theta)\left[\epsilon\,r\,J_{0}\left(\frac{\epsilon\,r}{v}\right)-2\,v\,J_{1}\left(\frac{\epsilon\,r}{v}\right)-\epsilon\,r\,J_{2}\left(\frac{\epsilon\,r}{v}\right)\right]}{4r}\right\} \,.\\
\bar{a}_{3}(\mathbf{r}) & =\frac{i\,(v+\delta v_{10})\left(\eta_{+}^{2}-\eta_{-}^{2}+\epsilon_{+}^{2}-\epsilon_{-}^{2}\right)\,\epsilon\cos(\theta)J_{1}\left(\frac{\epsilon\,r}{v}\right)}{4\,v^{2}(\eta_{+}+\eta_{-})}\,.
\end{align}
\\

\begin{align}
\bar{b}_{0}(\mathbf{r}) & =\frac{i\,\delta v_{20}\,\left(\eta_{+}^{2}+\eta_{-}^{2}+\epsilon_{+}^{2}+\epsilon_{-}^{2}\right)\,\epsilon\,\sin(\theta)\,J_{1}\left(\frac{\epsilon\,r}{v}\right)}{4\,v^{2}\,(\eta_{+}+\eta_{-})}\,.\label{eq:bbcoeff}\\
\bar{b}_{1}(\mathbf{r}) & =\frac{(v+\delta v_{10})(\eta_{+}\,\epsilon_{-}-\eta_{-}\,\epsilon_{+})-\delta v_{20}(\eta_{+}\,\eta_{-}+\epsilon_{+}\,\epsilon_{-})}{2\,v^{2}(\eta_{+}+\eta_{-})}\left\{ \frac{\sin(2\theta)\left[\epsilon\,r\,J_{0}\left(\frac{\epsilon\,r}{v}\right)-2\,v\,J_{1}\left(\frac{\epsilon\,r}{v}\right)-\epsilon\,rJ_{2}\left(\frac{\epsilon\,r}{v}\right)\right]}{4r}\right\} \,.\\
\bar{b}_{2}(\mathbf{r}) & =\frac{\delta v_{20}(\eta_{+}\,\eta_{-}+\epsilon_{+}\,\epsilon_{-})+(v+\delta v_{10})(\eta_{+}\,\epsilon_{-}-\eta_{-}\,\epsilon_{+})}{4\epsilon\,v^{2}\,(\eta_{+}+\eta_{-})}\Big\{\frac{\epsilon\,\cos(2\theta)\left(\epsilon\,r\,J_{0}\left(\frac{\epsilon\,r}{v}\right)-2\,v\,J_{1}\left(\frac{\epsilon\,r}{v}\right)-\epsilon\,r\,J_{2}\left(\frac{\epsilon\,r}{v}\right)\right)}{2r}\\
 & \times\,\epsilon^{2}\,J_{0}\left(\frac{\epsilon\,r}{v}\right)\Big\}\,.\nonumber \\
\bar{b}_{3}(\mathbf{r}) & =\frac{i\,\delta v_{20}\,\left(\eta_{+}^{2}-\eta_{-}^{2}+\epsilon_{+}^{2}-\epsilon_{-}^{2}\right)\,\epsilon\,\sin(\theta)\,J_{1}\left(\frac{\epsilon\,r}{v}\right)}{4\,v^{2}\,(\eta_{+}+\eta_{-})}\,.
\end{align}
\end{widetext}

\end{document}